\documentclass[journal]{IEEEtran}
\usepackage{subfigure}
\usepackage{colortbl}
\usepackage{bm}

\usepackage{graphicx}  
\usepackage{url}       

\usepackage{amsmath}   
\allowdisplaybreaks[4]
\usepackage{cite}

\usepackage{amsfonts,amssymb}

\usepackage{stfloats}

\usepackage{cases}
\usepackage{algorithm}
\usepackage{multirow}
\usepackage{algorithmic}
\usepackage{epstopdf}
\usepackage{color}

\usepackage[square, comma, sort&compress, numbers]{natbib}

\makeatletter
\newcommand{\vvast}{\bBigg@{3.0}}
\newcommand{\vast}{\bBigg@{4}}
\newcommand{\Vast}{\bBigg@{4.5}}
\newcommand{\VVast}{\bBigg@{5}}
\newcommand{\VVVast}{\bBigg@{5.5}}

\makeatother

\newtheorem{theorem}{{Theorem}}

\newcommand{\ls}[1]
{\dimen0=\fontdimen6\the\font
	\lineskip=#1\dimen0
	\advance\lineskip.5\fontdimen5\the\font
	\advance\lineskip-\dimen0
	\lineskiplimit=.9\lineskip
	\baselineskip=\lineskip
	\advance\baselineskip\dimen0
	\normallineskip\lineskip
	\normallineskiplimit\lineskiplimit
	\normalbaselineskip\baselineskip
	\ignorespaces
}

\begin{document}

	
		\title{\ls{1.0}{Statistical Delay and Error-Rate Bounded QoS Provisioning for AoI-Driven 6G Satellite-Terrestrial Integrated Networks Using FBC}}

\author{\IEEEauthorblockN{Jingqing Wang,~\IEEEmembership{Member,~IEEE}, Wenchi Cheng,~\IEEEmembership{Senior Member,~IEEE}, and H. Vincent Poor,~\IEEEmembership{Life Fellow,~IEEE}} \\[0.2cm]
	
	\thanks{
	This work was supported in part by National Key R\&D Program of China under Grant 2021YFC3002100.}
	\thanks{Jingqing Wang and Wenchi Cheng are with the State Key Laboratory of Integrated Services Networks, Xidian University, Xi’an, China (e-mails: wangjingqing00@gmail.com; wccheng@xidian.edu.cn).}
	\thanks{H. V. Poor is with the Department of Electrical and Computer Engineering, Princeton
		University, Princeton, NJ 08544 USA (e-mail: poor@princeton.edu).}
}

\maketitle


\begin{abstract}
As one of the pivotal enablers for 6G, satellite-terrestrial integrated networks have emerged as a solution to provide extensive connectivity and comprehensive 3D coverage across the spatial-aerial-terrestrial domains to cater to the specific requirements of 6G massive ultra-reliable and low latency communications (mURLLC) applications, while upholding a diverse set of stringent quality-of-service (QoS) requirements. 
In the context of mURLLC satellite services, the concept of data freshness assumes paramount significance, as the use of outdated data may lead to unforeseeable or even catastrophic consequences.
To effectively gauge the degree of data freshness for satellite-terrestrial integrated communications, the notion of age of information (AoI) has recently emerged as a novel dimension of QoS metrics to support time-sensitive applications. 
Nonetheless, the research efforts directed towards defining novel diverse statistical QoS provisioning metrics, including AoI, delay, and reliability, while accommodating the dynamic and intricate nature of satellite-terrestrial integrated environments, are still in their infancy.
To overcome these problems, in this paper we develop analytical modeling formulations/frameworks for statistical QoS over 6G satellite-terrestrial integrated networks using hybrid automatic repeat request with incremental redundancy (HARQ-IR) in the finite blocklength regime.
In particular, first we design the satellite-terrestrial integrated wireless network architecture model and AoI metric model.
Second, we characterize the peak-AoI bounded QoS metric using HARQ-IR protocol.
Third, we develop a set of new fundamental statistical QoS metrics in the finite blocklength regime.
Finally, extensive simulations have been conducted to assess and analyze the efficacy of statistical QoS schemes for satellite-terrestrial integrated networks.
\end{abstract}


\section{Introduction}\label{sec:intro}

In light of the global deployment of 5G wireless networks, researchers are actively engaged in conceptualizing the forthcoming 6G wireless networks~\cite{dang2020should} to cater to a wide range of emerging wireless applications that demand increasingly rigorous and varied quality-of-service (QoS) standards.
Providing distinct levels of QoS with specific delay constraints for diverse categories of time-sensitive wireless multimedia data constitutes a pivotal element within the framework of 6G mobile wireless networks.  
Consequently, the effective establishment of delay-bound QoS provisioning for real-time wireless services has garnered considerable research interest in recent years~\cite{wen2018,9130689}.
The inherently unstable characteristic of wireless channels and intricate, diverse, and ever-changing structures of 6G mobile wireless networks make it challenging to uphold the conventional deterministic QoS criteria for upcoming time and error-sensitive multimedia data transmission.
Thus, \textit{statistical QoS provisioning theory} has emerged as a powerful strategy for implementing delay-bounded QoS assurances in real-time communications.
However, the escalating volume of time/error-sensitive multimedia traffic mandates that 6G wireless networks address an array of intricate and demanding QoS prerequisites, including stringent end-to-end delay, ultra-high reliability, unmatched data rates, and others~\cite{10206807}.

One of the key focuses in developing 6G technology is the establishment of systems that facilitate the realization of \textit{massive Ultra-Reliable Low-Latency Communications} (mURLLC)~\cite{9311792}, demanding rigorous QoS guarantees to ensure their prompt and reliable delivery.
However, the significant challenge posed by the need for extensive connectivity and comprehensive coverage represents a pivotal obstacle when it comes to supporting mURLLC.
Satellite communication systems have been devised as a potential alternative to terrestrial communication systems to offer the promise of global coverage and seamless connectivity, while upholding stringent QoS standards. Conversely, conventional terrestrial networks still retain their pivotal role in furnishing cost-effective and high-speed wireless services, especially in densely populated urban areas.  Therefore, the integration of satellite and terrestrial networks presents an opportunity to harness the respective advantages of both systems, thereby enabling the provision of ubiquitous network services while supporting mURLLC.

To facilitate comprehensive 3D coverage for 6G services, employing the integration of space, aerial, and terrestrial components, satellite-terrestrial integrated networks~\cite{8894851} present significant potential in offering global coverage and seamless connectivity while facilitating various emerging time-sensitive applications. 
For these real-time applications, ensuring a variety of stringent QoS benchmarks becomes of paramount importance, as outdated and unreliable data may engender unpredictable or catastrophic outcomes.
Nevertheless, the complicated and dynamically evolving nature of satellite-terrestrial integrated network environments inevitably engenders system performance requirements, thus, in turn, would significantly impact QoS assurances in terms of data freshness, delay, and reliability.	
In light of the aforementioned challenges, it is imperative to develop potential diverse QoS measurement and control strategies, guaranteeing the stringent mURLLC requirements over 6G satellite-terrestrial integrated networks.

To effectively measure data freshness, recently, \textit{age of information} (AoI)~\cite{9109636,yates2021age,10098731} has surfaced as a novel QoS metric for accurately reflecting data packets' update speed in short-pack-based status update systems for delay/age-sensitive data transmissions, particularly for satellite applications~\cite{10206807}.
Due to the limited number of information bits in satellite status updates and the need for prompt delivery to remote destinations, long codewords impose significant demands on receiver's storage and computational resources, thereby diminishing the system's timeliness.
Previous research has introduced and examined techniques for small-packet communications, including finite blocklength coding (FBC)~\cite{yury2010,Yp2011,dosti2019performance}, as a means to alleviate access latency and decoding complexity while measuring/controlling the AoI metric across satellite-terrestrial integrated networks. 
This is particularly crucial in meeting the stringent QoS requirements of mURLLC.

Nevertheless, the complicated and dynamic nature of satellite-terrestrial integrated network environments makes transmission error an inevitable occurrence, thus, in turn, significantly {impacting} the system reliability.
To ensure the ultra-high reliability requirements for 6G satellite-terrestrial integrated communications, hybrid automatic repeat request (HARQ) protocols~\cite{6490421} have been implemented to adaptively manage data rate and increase transmission reliability.
The authors of~\cite{8525430} have presented a rapid HARQ protocol for enhancing the end-to-end throughput to support delay-constrained applications.
For satellite-terrestrial integrated communications, where end-to-end delay is non-negligible, the presence of substantial one-way delays can significantly hinder the effectiveness of real-time services.
In response to the above challenges, the authors of~\cite{9686609} investigate the efficacy of an age-oriented HARQ protocol, examining aspects such as decoding error probabilities and expected delay.


Despite the diligent efforts from both academia and industry for guaranteeing mURLLC services via satellite-terrestrial integrated networks, most previous studies have focused on investigating QoS metrics in terms of the delay-bound violating probability without considering the non-vanishing decoding error probability and the AoI violation probability. It is crucial to design and model satellite-terrestrial integrated wireless network architectures while taking into account stringent and diverse age/delay/error-rate bounded QoS guarantees by defining more fundamental statistical QoS performance metrics and characterizing their analytical relationships, such as delay-bound-violation probability, decoding error probability, and AoI violation probability, in the finite blocklength regime.
However, due to the lack of comprehensive understanding of QoS-driven fundamental modeling analyses over satellite-terrestrial integrated networks, how to efficiently identify and define novel FBC-based QoS metrics bounded by \textit{AoI}, \textit{delay}, and \textit{error-rate} as well as the associated relationships, including diverse \textit{QoS-exponent functions}, is still a challenging task.


To remedy the above deficiencies, we develop a series of fundamental statistical QoS metrics and their modeling/controlling techniques over 6G satellite-terrestrial integrated networks.
In particular, first we design the satellite-terrestrial integrated network architecture model and the AoI metric model.
Second, we derive the peak-AoI bounded QoS metric and the corresponding violation probability using HARQ-IR protocol.
Third, focusing on modeling and analyzing the fundamental performance, we develop new QoS metrics in terms of delay and error-rate in the finite blocklength regime.
Finally, we conduct extensive simulations to validate, evaluate, and analyze our developed statistical QoS schemes over satellite-terrestrial integrated 6G wireless networks.

The rest of this paper is organized as follows. Section~\ref{sec:sys} establishes the satellite-terrestrial integrated wireless communication models.
Section~\ref{sec:pAoI} analytically characterizes the peak-AoI bounded QoS metric using HARQ-IR.
Section~\ref{sec:EC1} develops statistical delay and error-rate bounded QoS performance metrics using FBC.
Section~\ref{sec:results} validates and evaluates our developed performance modeling techniques. The paper concludes with Section~\ref{sec:conclusion}.

	\section{The System Models}\label{sec:sys}
\begin{figure*}[!t]
	\vspace{-5pt}
	\centering
	\includegraphics[scale=0.41]{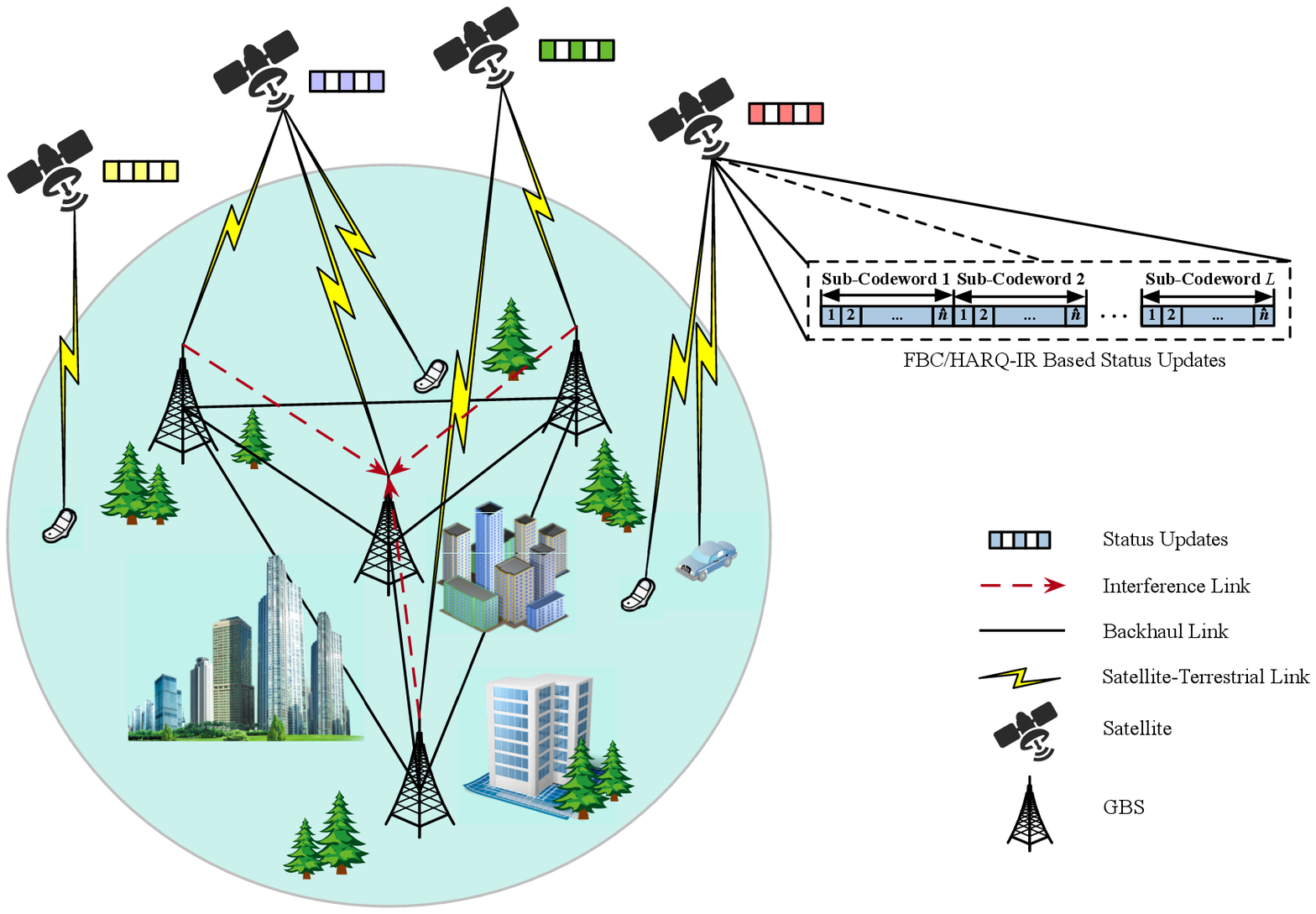}
	\caption{The system model for the satellite-terrestrial integrated wireless networks using HARQ-IR in the finite blocklength regime.}
	\label{fig:1}
	\vspace{-5pt}
\end{figure*}

Figure~\ref{fig:1} presents the downlink satellite-terrestrial integrated system architecture model for remote status delivery scenarios. 
The traditional Poisson point process (PPP) model can be used to characterize the locations of the base stations in cellular networks due to its inherent analytical tractability.
However, the spatial distribution of {ground base stations (GBSs)} follows more intricate patterns that present challenges for analytical modeling. These patterns, in general, deviate from the simplicity of the PPP.
In the context of a practical scenario, we characterize the spatial distribution of interfering GBSs using the Mat\'{e}rn hard-core point process (MHCPP) model that imposes a certain minimum distance, referred to as the hard-core distance, {ensuring} that neighboring GBSs are maintained at a certain distance from each other, which is more accurate and realistic than the PPP for modeling the spatial structure of GBSs.
The MHCPP model is characterized by an intensity denoted as $\lambda_{M}$ and a minimum distance of $d_{M}$ between distinct GBSs.
 
Our focus is directed towards the task of downlink transmission of remote sensing status updates, where the satellite is tasked with monitoring dynamic physical processes, which may include gathering vital signs data from astronauts or tracking the spacecraft's attitude, among other parameters. 
The satellite captures these status update messages and subsequently transmits them to the GBS, while ensuring stringent QoS requirements for mURLLC.
Without loss of generality, we assume that the GBSs (terrestrial interferers) are located in a finite annular area centered around the destination GBS with inner radius, denoted by $R_{\text{in}}$, and an outer radius, denoted by $R_{\text{out}}$, beyond which the interference is assumed to be negligible due to path loss~\cite{7997291}.
The proposed system model includes $N$ status-update data packets.
The index set, denoted by  $\textsf{N}$, represents all status-updates with a cardinality of $N$.
Consequently, by implementing FBC techniques, each status update packet $u$ $(u\in\textsf{N})$ transmitted from the satellite is encoded into a codeword employing $n$ channel uses.

\subsection{The Wireless Channel Model Over Satellite-Terrestrial Integrated Networks}
Define $h_{s}(u)$ as the channel fading coefficient during the transmission of status-update packet $u$ between the satellite and the destination GBS.
We assume that $h_{s}(u)$ follows shadowed-Rician distribution~\cite{1198102}, which can efficiently describe the channel characteristics over wireless land mobile satellite communication systems.  
Two statistical models delineating the land mobile satellite link are presented in~\cite{1623307} and~\cite{1198102}. In~\cite{1623307}, Rayleigh and lognormal distributions describe the amplitudes of the scatter and the line-of-sight (LOS) components, respectively. 
However, the derived expressions in~\cite{1623307} are notably intricate. 
Conversely,~\cite{1198102} employs the Nakagami-$m$ distribution to model the LOS amplitude instead of the lognormal distribution, demonstrating a favorable fit to experimental data for land mobile satellite links. Consequently, by adopting the model elucidated in~\cite{1198102}, the composite shadowed-Rician distribution can be accurately approximated through a distribution derived from the Rayleigh and Nakagami-$m$ distributions.
{Correspondingly, we first write the probability density function (PDF), denoted by $f_{|h_{s}(u)|^{2}}(x)$, of the channel gain $|h_{s}(u)|^{2}$ as follows~\cite{1198102}:
	\begin{equation}\label{equation_pdf}
		f_{|h_{s}(u)|^{2}}(x)=\alpha_{s}(u) e^{-\beta_{s}(u) x}  \sideset{_1}{_{1}}{\mathop{F}}(m_{s}(u),1,\delta_{s}(u) x), \,\, x>0,
	\end{equation}
	where
	\begin{equation}\label{equation02}
		\begin{cases}
			\alpha_{s}(u)=\frac{1}{2b_{s}(u)}\left[\frac{2b_{s}(u)m_{s}(u)}{2b_{s}(u)m_{s}(u)+\Omega_{s}(u)}\right]^{m_{s}(u)};\\
			\beta_{s}(u) = \frac{1}{2b_{s}(u)};\\
			\delta_{s}(u)= \frac{\Omega_{s}(u)}{2b_{s}(u)\left[2b_{s}(u)m_{s}(u)+\Omega_{s}(u)\right]},
		\end{cases}
	\end{equation}
	where $\Omega_{s}(u)$ denotes the average power of LOS component, $2b_{s}(u)$ is the average power of the multipath component, $m_{s}(u)\in[0,\infty]$ represents the Nakagami-$m$ parameter, and $_{1}F_{1}(\cdot,\cdot,\cdot)$ is the confluent hypergeometric function.
	When $m_{s}(u) = 0$, the shadowed-Rician fading reduces to Rayleigh fading. On the other hand, when $m_{s}(u) = \infty$, it converges to Rician fading.
	For simplicity, we suppose that the Nakagami-$m$ parameter $m_{s}(u)$ takes on integer values~\cite{6918460}.
	Correspondingly, we can obtain
	\begin{align}
		&\sideset{_1}{_{1}}{\mathop{F}}(m_{s}(u),1,\delta_{s}(u) x)
		\nonumber\\
		&\quad=e^{\delta_{s}(u) x}\sum_{l=0}^{m_{s}(u)-1}\frac{(-1)^{l}\left(1-m_{s}(u)\right)_{l}\left[\delta_{s}(u) x\right]^{l}}{(l!)^{2}}
	\end{align}
	where $(\cdot)_{l}$ denotes the Pochhammer symbol~\cite{LS2007}.}

The signal to interference and noise ratio (SINR), denoted by $\gamma_{s}(u)$, when transmitting status-update data packet $u$ from the satellite to the destination GBS is derived as follows:
\begin{equation}\label{equation04}
	\gamma_{s}(u)=\frac{\phi_{s}(u){\cal P}_{s}(u)|h_{s}(u)|^{2}}{I_{\text{a}}+1}
\end{equation}
where ${\cal P}_{s}(u)=P_{s}(u)/\sigma^{2}$ represents the transmit signal-to-noise ratio (SNR) of the satellite, where $P_{s}(u)$ is the transmit power at the satellite and $\sigma^{2}$ is the noise power, $\phi_{s}(u)$ and $\phi_{j}(u)$ are the link responses between the satellite and the destination GBS and between GBS $j$ and the destination GBS, respectively.
By considering free-space path loss and antenna gain, the link responses can be modeled as follows:
\begin{equation}\label{equation05}
	\begin{cases}
			\phi_{s}(u)\triangleq\left(\frac{c}{4\pi f_{c}d_{s}(u)}\right)^{2}G_{s}G_{d}(u);\\
			\phi_{j}(u)\triangleq\left(\frac{c}{4\pi f_{c}d_{j}(u)}\right)^{2}G_{j}(u)G_{d}(u),
	\end{cases}
\end{equation}
where $c$ denotes the speed of light, $f_{c}$ is the frequency, $d_{s}(u)$ and $d_{j}(u)$ denote the distances when transmitting status-update data packet $u$ between the satellite and the destination GBS and between GBS $j$ and the destination GBS, respectively, $G_{s}$ is the antenna gain at the satellite, $G_{d}(u)$ and $G_{j}(u)$ are the antenna gains at the destination GBS and GBS $j$, respectively, and $I_{\text{a}}$ is the aggregate interference power received from terrestrial interferers (nearby GBSs) with the transmit power, denoted by $P_{t}(u)$, which is given as follows: 
\begin{equation}
	I_{\text{a}}=\sum\limits_{j=1}^{K}\phi_{j}(u)P_{t}(u)|h_{j}(u)|^{2}[d_{j}(u)]^{-\widetilde{\alpha}}
\end{equation}
where $K$ is the number of interfering GBSs, ${\cal P}_{t}(u)=P_{t}(u)/\sigma^{2}$ represents the transmit SNR of the terrestrial interferers, $h_{j}(u)$ is the terrestrial channel fading coefficient from the GBS $j$ to the destination GBS, which follows Rayleigh distribution with scale parameter $\varsigma$, and $\widetilde{\alpha}$ is the path loss exponent.

\subsection{The Maximum Achievable Coding Rate Within FBC Regime}

Denote by $\epsilon_{s}(u)$ the decoding error probability for the transmission of status-update $u$ across our proposed satellite-terrestrial integrated wireless networks using FBC.

\textit{Definition 1. The $(n, M_{s}, \epsilon_{s}(u))$-Code:}
An $(n,M_{s}, \epsilon_{s}(u))$-code is defined as follows:

We consider a message $W_{s}$ uniformly distributed across the elements of ${\cal M}_{s}$, where ${\cal M}_{s}$ is a set of messages.
\begin{itemize}
	\item An encoder $\Upsilon$: $\{ 1,\dots,M_{s}\} \mapsto{\cal A}^{n}$ is tasked with the assignment of message $W_{s}$ to a codeword having a length of $n$. Here, ${\cal A}^{n}$ signifies the codebook, which encompasses the entire array of possible codewords generated by the encoding function denoted as $\Upsilon$.

	\item A decoder ${\cal D}$: ${\cal B}^{n}\mapsto\{1,\dots,M_{s}\}$ is tasked with the assignment of decoding received message, yielding $\widehat{W}_{s}$ as the estimated signal at the destination node, where ${\cal B}^{n}$ is the set of received codewords with length $n$.
	We define the upper-bounded decoding error probability as follows:
	\begin{equation}\label{equation007}
		\text{Pr}\left\{\widehat{W}_{s}\neq W_{s}\right\}= \epsilon_{s}(u).
	\end{equation}
\end{itemize}

Based on Shannon's capacity theorem, the channel capacity, a fundamental concept, represents the maximum rate at which information can be transmitted without regard to the desired error probability, provided the blocklength is allowed to grow without bound. In practical scenarios, a crucial consideration involves evaluating the deviation from channel capacity necessary to maintain a specified error probability at finite blocklength.
Shannon's capacity theory does not offer explicit guidance to address this inquiry, neither through the strong version of the coding theorem nor the reliability function, {denoted by $E(W_{s}, R_{s}(u))$, which quantifies the asymptotic behavior of the decoding error probability, i.e.,~\cite{1988Principles}
	\begin{equation}
		E(W_{s},R_{s}(u))=-\lim\limits_{n\rightarrow\infty}\sup \frac{1}{n}\log_{2} P_{e,\max}\left(W_{s},R_{s}(u), n\right)
	\end{equation}
where $R_{s}(u)$ is the transmission rate and $P_{e,\max}\left(W_{s},R_{s}(u), n\right)$ denotes the minimum error probability over all codes of block length $n$ and with message set ${\cal M}_{s}$.
As the transmission rate approaches capacity, the reliability function describes how rapidly the decoding error probability decays.}
To address this, subsequent to Shannon's establishment of the convergence of optimal coding rate to capacity, research endeavors have been undertaken to assess the penalty incurred by finite blocklength. Recent studies have introduced several new achievability and converse bounds that tightly constrain the fundamental limits for blocklengths as short as 100 by utilizing FBC.
In the context of supporting stringent and diverse QoS provisioning of mURLLC, it is crucial to formulate the coding rate model considering finite data-packet sizes.
As a result, we propose an alternative solution for providing statistical QoS guarantees through implementing FBC, which allows us to attain statistically assured mURLLC-QoS requirements while effectively managing small violation probabilities.
The decoding error probability $\epsilon_{s}(u)$ during the transmission of status update $u$ of the developed performance modeling frameworks is determined as follows~\cite{6802432}:
\begin{equation}\label{equation09}
	\epsilon_{s}(u)\approx \mathbb{E}_{\gamma_{s}(u)} \left[{\cal Q}\left(\frac{\sqrt{n}\left(C(\gamma_{s}(u))-R^{*}_{s}\right)}{\sqrt{V(\gamma_{s}(u))}}\right)\right]
\end{equation}
where $\mathbb{E}_{\gamma_{s}(u)}[\cdot]$ represents the expectation over the SINR $\gamma_{s}(u)$, ${\cal Q}(\cdot)$ represents the \textit{$Q$}-function, $R^{*}_{s}$ denotes the \textit{maximal achievable coding rate}, and $C(\gamma_{s}(u))$ and $V(\gamma_{s}(u))$ are the \textit{channel capacity} and \textit{channel dispersion}.
Specifically, the channel capacity $C(\gamma_{s}(u))$ follows the Shannon-Hartley theorem.
The channel dispersion $V(\gamma_{s}(u))$ (measured in squared information units per channel use) of a channel with channel capacity $C(\gamma_{s}(u))$,  which measures the stochastic variability of the channel relative to a deterministic channel with the same capacity, can be calculated as follows~\cite{yury2010}:
\begin{equation}
V(\gamma_{s}(u))\!=\!\!\!\lim\limits_{\epsilon_{s}(u)\rightarrow0}\lim\limits_{n\rightarrow\infty}\!\!\sup\left\{\frac{1}{n}\left[\frac{nC(\gamma_{s}(u))\!-\!\log (M_{s}^{*})}{{\cal Q}^{-1}(\epsilon_{s}(u))}\right]^{2}\right\}
\end{equation}
where $M_{s}^{*}$ is the maximum achievable codebook size and ${\cal Q}^{-1}(\cdot)$ represents the inverse \textit{$Q$}-function.
Based on the findings presented in~\cite{yury2010}, the authors have demonstrated the expressions for the channel capacity and channel dispersion functions as follows:
\begin{equation}
	\begin{cases}	\vspace{1pt}
		C(\gamma_{s}(u))
		=\log_{2}\left[1+\gamma_{s}(u)\right]; \\
		V(\gamma_{s}(u))=
		1-\frac{1}{\left[1+\gamma_{s}(u)\right]^{2}}.
	\end{cases}
\end{equation}

\subsection{The Peak AoI Metric Model Using HARQ-IR}
	\begin{figure}[!t]
		\vspace{-5pt}
	\centering
	\includegraphics[scale=0.34]{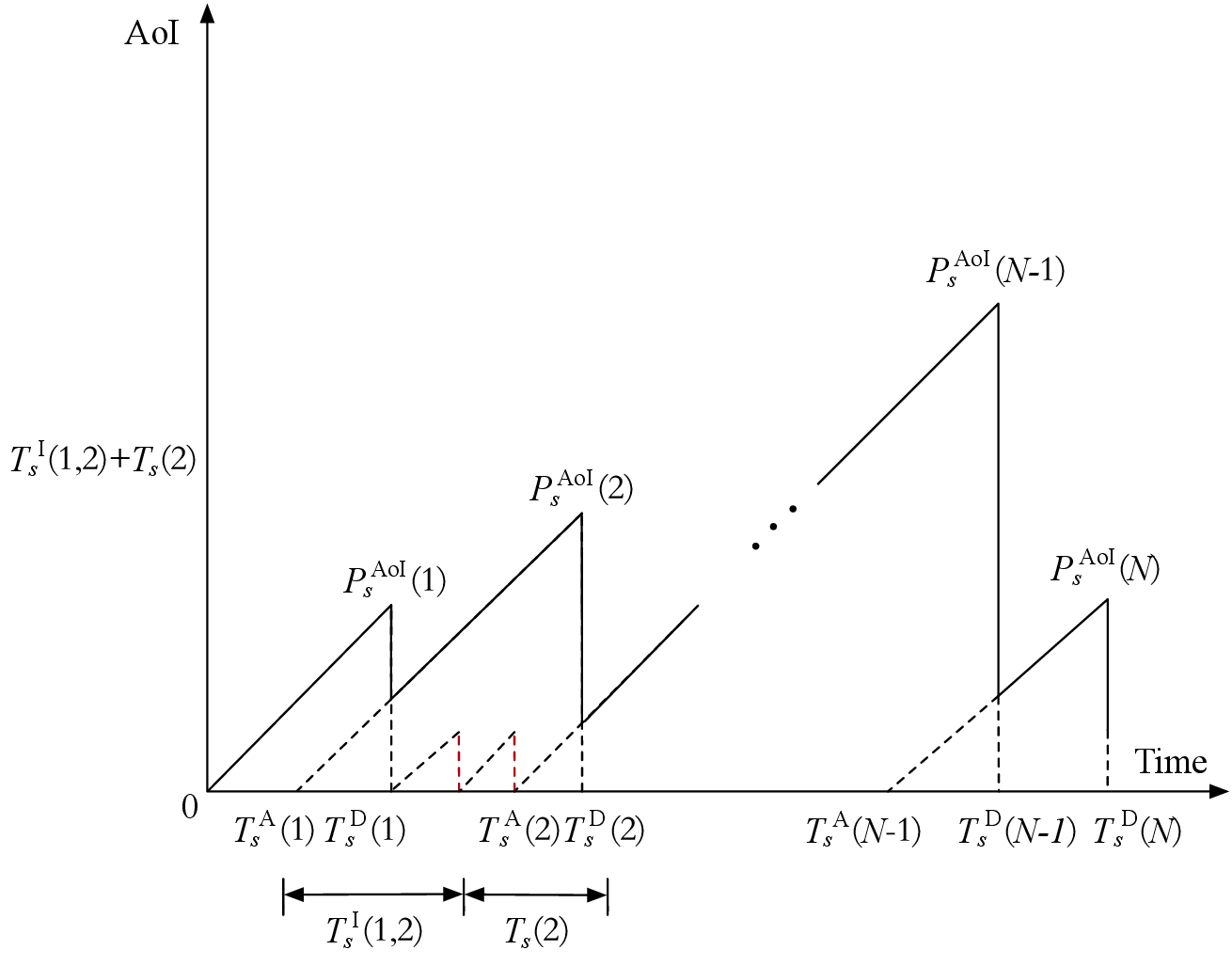}
	\caption{AoI evolution for $N$ finite-blocklength status update packets.}
	\label{fig:2}
	\vspace{-5pt}
\end{figure}
Traditionally, the HARQ has been proposed as a transmission scheme, aiming at ensuring reliability. This method incorporates forward error-correcting coding in the PHY and retransmissions in the data link/medium access layer using ACK/NACK feedback to enable reliable transmission {over} time-varying channel. 
However, in HARQ, unsuccessful transmissions result in the retransmission of the entire packet, contributing to a high-delay scheme. This delay stems from multiple decoding attempts and subsequent retransmissions.
In response to this, the HARQ-IR protocol has been developed to enable more bandwidth-efficient partial retransmissions, ensuring reliability without introducing excessive delay. 
With HARQ-IR, only the missing or incorrect segments of the data packet are retransmitted, constituting incremental redundancy.
Following the transmission of a sub-codeword, the source node pauses and awaits feedback signals to determine whether to proceed with the current transmission. Simultaneously, the decoder at the receiver node remains active, combining sub-codewords at the end of each round to decode the updates. Subsequently, instantaneous feedback signals are transmitted to the source node, communicating decoding results and requesting additional information to rectify errors in the received packet.
Thus, the HARQ-IR protocol proves more efficient in utilizing bandwidth and exhibits adaptability through dynamic adjustments of redundancy levels based on channel feedback.

By incorporating HARQ-IR protocol, each finite length-$n$ codeword is divided into $L$ blocks, each consisting of $\widehat{n}$ symbols and the data blocks are transmitted sequentially in consecutive time slots. 
The codeword, consisting of a series of data blocks with a finite length of $L$, is defined as $\bm{x}_{s}(u)\triangleq\left[\bm{x}_{s}^{(1)}(u),\dots,\bm{x}_{s}^{(L)}(u)\right]$, where  $\bm{x}_{s}^{(l)}(u)\triangleq\left[\bm{x}_{s}^{(\widehat{n}(l-1)+1)}(u),\dots, \bm{x}_{s}^{(\widehat{n}l)}(u)\right]$ for $l=1,\dots, L$.
Using HARQ-IR, in the event that the receiver successfully deciphers the received data packet, it will initiate the process of transmitting an ACK feedback signal, subsequently effecting the removal of this packet in the buffer.
Conversely, if the received packet cannot be successfully decoded, a NACK is sent back to the transmitter, prompting the transmission of another data block.
This process continues until either the codeword is successfully decoded at the receiver or the maximum number of allowed transmissions $L_{\max}$ for the packet is reached.
The initial coding rate, denoted by $R^{*}_{s,\text{in}}$, is derived as follows:
\begin{equation}\label{equation66}
	R^{*}_{s,\text{in}}\triangleq\frac{\log (M_{s}^{*})}{\widehat{n}}\,\,\, \text{bits/channel use}.
\end{equation}
The maximal achievable coding rate, denoted as $R^{*}_{s,l}$, after HARQ-IR retransmission round $l$ is defined as follows:
\begin{equation}\label{equation67}
	R^{*}_{s,l}\triangleq\frac{\log M_{s}^{*}}{\widehat{n}l}=\frac{R^{*}_{s,\text{in}}}{l} \,\,\, \text{bits/channel use}.
\end{equation}
Correspondingly, to measure data freshness, AoI is adopted as a fundamental QoS performance metric.
The relevant notations and definitions are illustrated in Fig.~\ref{fig:2}.
As shown in Fig.~\ref{fig:2}, there are two rounds of HARQ-IR retransmissions before achieving the successful transmission of the second status update packet.
Considering the finite-blocklength status update, denoted as $u$, we represent the arrival, service, and departure time as $T_{s}^{\text{A}}(u)$, $T_{s}^{\text{S}}(u)$, and $T_{s}^{\text{D}}(u)$ $(u\in\textsf{N})$, respectively.
For the sake of simplicity, we set the arrival time of the first data packet as $T_{s}^{\text{A}}(0) = 0$.
Furthermore, we introduce the following inter-arrival time between two status-updates $v$ and $u$ $(1\leq v\leq u)$:
\begin{equation}\label{equation003}
	T_{s}^{\text{I}}(v,u)=\sum_{j=v+1}^{u}T_{s}^{\text{I}}(j-1,j).
\end{equation}
We can derive the time for cumulative service, denoted by $T_{s}^{\text{S}}(v,u)$, when transmitting status update $v$ to status update $u$ as $T_{s}^{\text{S}}(v,u)=\sum_{j=v}^{u}T_{s}^{\text{S}}(j)$.
Then, the time of departure $T_{s}^{\text{D}}(u)$ for transmitting $u$ status-updates is derived as follows~\cite{chang2000performance}:
\begin{align}\label{equation015}
	T_{s}^{\text{D}}(u)=\sup_{v\in\textsf{N}, v\leq u}\left\{T_{s}^{\text{A}}(v)+T_{s}^{\text{S}}(v,u)\right\}.
\end{align}
Accordingly, the total sojourn time, denoted by $T_{s}(u)$, when transmitting status update $u$ is determined as follows:
\begin{equation}\label{equation005}
	T_{s}(u)\triangleq T_{s}^{\text{D}}(u)-T_{s}^{\text{A}}(u)
	=\sup_{v\in\textsf{N}, v\leq u}\left\{T_{s}^{\text{S}}(v,u)-T_{s}^{\text{I}}(v,u)\right\}.
\end{equation}
Fig.~\ref{fig:2} shows that the peak AoI, denoted as $PAoI_{s}(u)$, of our proposed satellite-terrestrial integrated system model during the transmission of status update $u$ is represented as $P_{s}^{\text{AoI}}(u)=T_{s}^{\text{I}}(u-1,u)+T_{s}(u)$.

\section{The Peak-AoI Bounded QoS Metric and Peak AoI Violation Probability Using HARQ-IR}\label{sec:pAoI}

One critical aspect to consider for designing satellite-terrestrial integrated wireless services is the efficient guarantee of low-tail requirements for age/time/reliability-sensitive data transmissions, which play a crucial role in the fundamental performance analysis and provide practical engineering guidance for designing, analyzing, and evaluating diverse statistical QoS provisioning schemes.
Diverse statistical QoS metrics can provide valuable insights into the worst-case scenario performances for our proposed schemes.
However, the analysis of tail behaviors, i.e., the peak AoI violation probability, presents several challenges.
As a result, our proposed modeling schemes prioritize facilitating QoS provisioning for low-tail cases. 
This emphasis is directed towards evaluating the violation probability, ensuring efficient support for the specific requirements associated with low-tail considerations for age-sensitive data transmissions, rather than solely assessing average performance metrics~\cite{9324753}.
However, traditional statistical QoS theory only investigates low-tail metrics in terms of queueing delay, without analyzing tail behaviors for AoI and error-rate. The reduction of violation probability that AoI of status updates exceeds a given age constraint is of great significance for guaranteeing the data freshness in our proposed systems.	
Consequently, this section aims to characterize the violation probabilities with regard to peak AoI through using HARQ-IR, which can severely impact network performance.

\subsection{The Peak-AoI Bounded QoS Metric}\label{sec:pAoI1}

Owing to the stochastic characteristics in both the arrival and service times, the direct determination of the peak AoI violation probability becomes impractical.
To address this challenge, we employ stochastic network calculus (SNC) to measure the stochastic arrival and service times.
Usually, the network-layer model with arrival, departure, and service processes is investigated as residing in a bit domain, where traffic and service {are} measured in bits. 
We can derive performance bounds in the SNR domain~\cite{HAL2016}, and then map the results to the bit domain to obtain network-layer bounds for peak AoI.
Observing that the bit and SNR domains are linked by the exponential function, we employ SNC to convert the stochastic arrival and service times into the SNR domain, which are represented as follows~\cite{HAL2016}:
	\begin{equation}
		\begin{cases}
			{\cal T}_{s}^{\text{I}}(v,u)\triangleq e^{T_{s}^{\text{I}}(v,u)}; \\
			{\cal T}_{s}^{\text{S}}(v,u)\triangleq e^{T_{s}^{\text{S}}(v,u)}.
		\end{cases}
	\end{equation}
In the SNR domain, the peak AoI is defined as ${\cal P}_{s}^{\text{AoI}}(u)={\cal T}_{s}^{\text{I}}(u-1,u){\cal T}_{s}(u)$, where ${\cal T}_{s}^{\text{I}}(u-1,u)$ and ${\cal T}_{s}(u)$ are the time of inter-arrival between status update $(u-1)$ and status update $u$ and total sojourn time, respectively, which are determined as follows:
	\begin{equation}\label{equation12}
		\begin{cases}
			{\cal T}_{s}^{\text{I}}(u-1,u)=e^{T_{s}^{\text{I}}(u-1,u)};\\ 
			{\cal T}_{s}(u)=e^{ T_{s}(u)}. 
		\end{cases}
	\end{equation}

In the SNR domain, our derivations necessitate not only the computation of products and quotients involving random variables, but also necessitates determining infimums and supremums.
To streamline these calculations, we employ the Mellin transform, mitigating what would otherwise be intricate computations~\cite{schiessl2015delay}. The Mellin transform of a non-negative random variable $\bm{X}$ is defined as ${\cal M}_{\bm{X}}(\theta) \triangleq \mathbb{E}\left[\bm{X}^{(\theta-1)}\right]$, where $\theta > 0$ serves as a free parameter. Subsequently, the computational challenge is simplified to determine the Mellin transform for arrival and service times.
As the direct computation of the peak AoI violation probability is infeasible, an upper-bound is established by employing the Mellin transform.
For the peak AoI in the SNR domain, the Mellin transform, denoted as ${\cal M}_{{\cal P}_{s}^{\text{AoI}}(u)}(\theta_{\text{AoI}})$, is obtained as follows:
	\begin{align}
		{\cal M}_{{\cal P}_{s}^{\text{AoI}}(u)}(\theta_{\text{AoI}})
	&={\cal M}_{{\cal T}_{s}^{\text{I}}(u-1,u)}(\theta_{\text{AoI}}){\cal M}_{{\cal T}_{s}(u)}(\theta_{\text{AoI}})
	\end{align}
where 
${\cal M}_{{\cal T}{s}^{\text{I}}(u-1,u)}(\theta_{\text{AoI}})$ is the Mellin transform of the inter-arrival time between status updates $(u-1)$ and $u$ in the SNR domain, while ${\cal M}_{{\cal T}{s}(u)}(\theta_{\text{AoI}})$ denotes the Mellin transform in terms of the sojourn time of status update $u$.

Accordingly, to support the peak-AoI bounded QoS provisioning, we propose to characterize the probability of peak AoI violation.
The peak AoI violation probability, denoted as $p_{s}^{\text{AoI}}(u)$, for transmitting status update $u$, can be determined by the probability that the peak AoI exceeds the designated peak AoI threshold, i.e.,
\begin{align}\label{equation49}
	p_{s}^{\text{AoI}}(u)\triangleq\text{Pr}\left\{ P_{s}^{\text{AoI}}(u)>\frac{A_{\text{th}}}{n}\right\}
\end{align}
where $A_{\text{th}}$ is the peak AoI violation threshold in channel uses through using FBC.
A more subtle question is how to obtain the tight bounds on the peak AoI violation probability in terms of the distribution of a random variable based on its Mellin transform. Accordingly, we proceed to obtain the upper-bounded peak AoI violation probability, which is given in the subsequent Lemma.

 \textit{Lemma 1}:
		The upper-bound on the peak AoI violation probability $p_{s}^{\text{AoI}}(u)$ over the developed satellite-terrestrial integrated wireless networks for a given the peak AoI threshold $A_{\text{th}}$ is determined as follows:
		\begin{align}\label{equation026a}
			p_{s}^{\text{AoI}}(u)&\leq e^{- \frac{\theta_{\text{AoI}} A_{\text{th}}}{n}}\mathsf{K}(\theta_{\text{AoI}},u)
		\end{align}
		where
		\begin{align}\label{equation027a}
			\mathsf{K}_{s}(\theta_{\text{AoI}},u)\triangleq&{\cal M}_{{\cal T}_{s}^{\text{I}}(u-1,u)}(1+\theta_{\text{AoI}})
			\Bigg[\sum_{v=1}^{u} {\cal M}_{{\cal T}_{s}^{\text{S}}(v,u)}(1+\theta_{\text{AoI}})
			\nonumber\\
			&\times
		 {\cal M}_{{\cal T}_{s}^{\text{I}}(v,u)}(1-\theta_{\text{AoI}})\Bigg]
		\end{align}
where $\theta_{\text{AoI}}>0$ is defined as the as the peak-AoI bounded QoS exponent to describe the quantify the rate of exponential decay of the QoS violation probability in terms of the peak AoI violation probability within the framework of our proposed modeling schemes for satellite-terrestrial integration.
Specifically, the peak-AoI bounded QoS exponent is an indicator of the level of stringency associated with statistical peak-AoI bounded QoS provisioning as the threshold increases, which describes the relationship between the peak AoI threshold and the probability of peak AoI exceeding that threshold.

	\begin{IEEEproof}
			The proof of Lemma~1 is in Appendix A.
	\end{IEEEproof}
	
	\indent\textit{Remarks on Lemma~1}: Lemma~1 provides a comprehensive characterization of the probability of upper-bounded peak AoI violations in the context of our developed satellite-terrestrial integrated modeling frameworks.
	This result facilitates a meaningful comparison with the delay violation probability, enabling the assurance of diverse statistical QoS schemes.
	
	Based on the results in Lemma~1, we can formally define the peak-AoI bounded QoS exponent as follows.
	
	\textit{Definition 2: The peak-AoI bounded QoS exponent:} By using the \textit{large deviations principle} (LDP)~\cite{ellis2006entropy}, the peak-AoI bounded QoS exponent $\theta_{\text{AoI}}$ quantifies the rate of exponential decay of the QoS violation probability in terms of the peak AoI violation probability within the framework of our proposed modeling schemes for satellite-terrestrial integration.
	
\indent\textit{Remarks on Definition~2}: Real-time multimedia services necessitate constraints on age, delay, error rate, or equivalently, a guaranteed effective bandwidth. 
In the context of these services, once a received real-time packet surpasses its specified bound, it is deemed useless and subsequently discarded. Nevertheless, achieving a hard delay-bound guarantee becomes practically infeasible over mobile wireless networks, primarily due to the influence of time-varying fading across wireless channels.
In light of this challenge, we explore an alternative approach by ensuring a peak AoI bound with a minimal probability of violation.
Accordingly, our proposed strategy aims to alleviate the tail behaviors of the peak AoI. 
This peak-AoI violation probability quantitatively represents the extreme behaviors of the peak AoI, which can then be used to compare with the delay and error-rate violation probabilities. 
By extending/applying the theories of age/delay/error-rate bounded statistical QoS, it serves as a bridge connecting PHY layer to the upper layer protocols.

	\subsection{The Mellin Transform of Inter-Arrival and Service Times}
Throughout the paper, we generally assume that the arrival and service processes are independent of each other. Considering general arrival and service processes, the server is a G$|$G$|$1 queue. 
Accordingly, the findings provide us with the ability to extend recent conclusions regarding the rate of tail decay obtained for \textit{independent and identically distributed} (i.i.d.) and service time, specifically within a General Independent (GI) service model~\cite{8242677}.

We consider the arrival and service processes falling within the broad class of $(\sigma,\rho)$-constrained process~\cite{chang2000performance}. The parameters $(\sigma,\rho)$ specify an affine bounding function, i.e., intercept and slope, of the logarithm of the moment generating function (MGF), and can be thought of as a stochastic version of the burst and rate parameters of a leaky-bucket regulator. Accordingly, we adopt the $(\sigma,\rho)$ constraint to our proposed system systems.
	
	\subsubsection{The Mellin Transform of the Inter-Arrival Time}
	Utilizing the $(\sigma_{s}^{\text{I}}(\theta_{\text{AoI}}),\rho_{s}^{\text{I}}(\theta_{\text{AoI}}))$-bounded process~\cite{chang2000performance}, we can establish an upper-bound expression for the Mellin transform in terms of the inter-arrival time in the SNR domain as follows:
	\begin{align}\label{equation18}
		{\cal M}_{{\cal T}_{s}^{\text{I}}(v,u)}(1+\theta_{\text{AoI}})
		&\leq e^{\theta_{\text{AoI}}\left[(u-v)\rho_{s}^{\text{I}}\left(\theta_{\text{AoI}}\right)+\sigma_{s}^{\text{I}}\left(\theta_{\text{AoI}}\right)\right]}.
	\end{align}
	 Setting $v=(u-1)$, the Mellin transform of the inter-arrival time in the SNR domain is derived as follows:
	\begin{equation}\label{equation19}
		{\cal M}_{{\cal T}_{s}^{\text{I}}(u-1,u)}(1+\theta_{\text{AoI}})\leq e^{\theta_{\text{AoI}}\left[\rho_{s}^{\text{I}}\left(\theta_{\text{AoI}}\right)+\sigma_{s}^{\text{I}}\left(\theta_{\text{AoI}}\right)\right]}.
	\end{equation}
Specifically, considering the special case of GI arrival processes, i.e., i.i.d. inter-arrival times $T_{s}^{\text{I}}(u-1,u)$, 
 we can obtain
	\begin{align}\label{equation19b}
		{\cal M}_{{\cal T}_{s}^{\text{I}}(v,u)}\!(1\!+\!\theta_{\text{AoI}})&=\mathbb{E}\!\left[\prod_{u=v+1}^{u}\!\!\!\!e^{\theta_{\text{AoI}} T_{s}^{\text{I}}(u-1,u)}\!\right]
			\nonumber\\
		&=\left[{\cal M}_{{\cal T}_{s}^{\text{I}}(u-1,u)}(1+\theta_{\text{AoI}})\right]^{(u-v)}.
	\end{align}	
	
	\subsubsection{The Mellin Transform of the Service Time}
		Utilizing the $(\sigma_{s}^{\text{S}}(\theta_{\text{AoI}}),\rho_{s}^{\text{S}}(\theta_{\text{AoI}}))$-bounded process, we can establish an upper-bound expression for the Mellin transform of the cumulative service time in the SNR domain as follows:
	\begin{align}\label{equation18b}
		{\cal M}_{{\cal T}_{s}^{\text{S}}(v,u)}(1+\theta_{\text{AoI}})&=\mathbb{E}\left[\left\{{\cal T}_{s}^{\text{S}}(v,u)\right\}^{\theta_{\text{AoI}}}\right]
			\nonumber\\
		&\leq e^{\theta_{\text{AoI}}\left[(u-v+1)\rho_{s}^{\text{S}}\left(\theta_{\text{AoI}}\right)+\sigma_{s}^{\text{S}}\left(\theta_{\text{AoI}}\right)\right]}.
	\end{align}
Specifically, considering the special case of GI service times, i.e., i.i.d. service times for every status-update packet, in the SNR domain the Mellin transform of the service time is obtained by ${\cal M}_{{\cal T}_{s}^{\text{S}}(u)}(1+\theta_{\text{AoI}})={\cal M}_{{\cal T}_{s}^{\text{S}}(1)}(1+\theta_{\text{AoI}})$, $\forall u$.
	Therefore, in the SNR domain, the Mellin transform of the cumulative service time is obtained as follows:
	\begin{align}\label{equation18d}
		{\cal M}_{{\cal T}_{s}^{\text{S}}(v,u)}(1+\theta_{\text{AoI}})&=\mathbb{E}\left[\,\prod_{u=v+1}^{u}e^{\theta_{\text{AoI}} T_{s}^{\text{S}}(u)}\right]
		\nonumber\\
		&
		=\left[{\cal M}_{{\cal T}_{s}^{\text{S}}(u)}(1+\theta_{\text{AoI}})\right]^{(u-v)}.
	\end{align}	
Drawing upon the assumption of i.i.d. inter-arrival and service times, we establish the upper bound for the probability of peak AoI violation, which is formulated in the subsequent theorem.

	\begin{theorem}\label{theorem002a}
	\underline{Claim 1.}  Considering general arrival and service processes, the upper bound on the peak AoI violation probability for the developed performance modeling frameworks is determined as follows:
		\begin{equation}\label{equation17a}
			p_{s}^{\text{AoI}}(u)\leq \xi e^{-\frac{\theta_{\text{AoI}} A_{\text{th}}}{n}}
	\end{equation}
where
\begin{equation}\label{equation034}
\xi	\triangleq\frac{e^{\theta_{\text{AoI}}\left[\rho_{s}^{\text{I}}\left(\theta_{\text{AoI}}\right)+\sigma_{s}^{\text{I}}\left(\theta_{\text{AoI}}\right)\!\right]}
		e^{\theta_{\text{AoI}}\left[\sigma_{s}^{\text{I}}\left(\!-\theta_{\text{AoI}}\right)+\rho_{s}^{\text{S}}\left(\theta_{\text{AoI}}\right)+\sigma_{s}^{\text{S}}\left(\theta_{\text{AoI}}\right)\!\right]}}
	{1-e^{-\theta_{\text{AoI}}\left[\rho_{s}^{\text{I}}\left(-\theta_{\text{AoI}}\right)-\rho_{s}^{\text{S}}\left(\theta_{\text{AoI}}\right)\right]}}
\end{equation}
while the stability condition  $\rho_{s}^{\text{S}}\left(\theta_{\text{AoI}}\right)<\rho_{s}^{\text{I}}\left(-\theta_{\text{AoI}}\right)$ holds.

\underline{Claim 2.} In the special case of GI$|$GI arrival and service processes, the upper bound on the peak AoI violation probability for the developed performance modeling frameworks is determined as follows:
		\begin{equation}\label{equation17}
			p_{s}^{\text{AoI}}(u)\leq \frac{e^{-\frac{\theta_{\text{AoI}} A_{\text{th}}}{n}}{\cal M}_{{\cal T}_{s}^{\text{I}}(u-1,u)}(1+\theta_{\text{AoI}}){\cal M}_{{\cal T}_{s}^{\text{S}}(u)}(1+\theta_{\text{AoI}})}{1-{\cal M}_{{\cal T}_{s}^{\text{I}}(u-1,u)}(1-\theta_{\text{AoI}}){\cal M}_{{\cal T}_{s}^{\text{S}}(u)}(1+\theta_{\text{AoI}})}
		\end{equation}
		while the stability condition $
		{\cal M}_{{\cal T}_{s}^{\text{I}}(u-1,u)}(1-\theta_{\text{AoI}}){\cal M}_{{\cal T}_{s}^{\text{S}}(u)}(1+\theta_{\text{AoI}})<1$ holds.
	\end{theorem}
	\begin{IEEEproof}
				The proof of Theorem~\ref{theorem002a} is in Appendix B.
	\end{IEEEproof}
	
	\indent\textit{Remarks on Theorem~\ref{theorem002a}:} Theorem~\ref{theorem002a} delves into fundamental analytical modeling techniques to quantify data freshness by deriving the peak AoI violation probability.
	Such fundamental performance analyses provide invaluable engineering guidance for the design, analysis, and evaluation of our proposed performance modeling schemes to ensure diverse statistical QoS provisioning, contributing to their practical utility in real-world applications.
	A significant characteristic of both the G$|$G and GI$|$GI results is its demonstration that the peak AoI violation probability exhibits an exponential tail decay represented as $\xi e^{-\frac{\theta_{\text{AoI}} A_{\text{th}}}{n}}$ with the delay rate $\theta_{\text{AoI}}$.
	In addition, assuming the i.i.d. exponential inter-arrival and service times, the results in Theorem 1 can be extended into the M$|$M$|$ 1queue.
		
	Furthermore, assuming that the arrival updates follow Poisson process characterized by a rate of $\lambda_{s}$, the inter-arrival time is modeled as an exponential process with a rate parameter $\lambda_{s}$.
	Accordingly, we have
	\begin{equation}\label{equation32}
		\rho_{s}^{\text{I}}\left(\theta_{\text{AoI}}\right)=\frac{1}{\theta_{\text{AoI}}}\log\left(\frac{\lambda_{s}}{\lambda_{s}-\theta_{\text{AoI}}}\right).
	\end{equation}
	The Mellin transform of the inter-arrival time is expressed as follows:
	\begin{equation}\label{equation33}
		{\cal M}_{{\cal T}_{s}^{\text{I}}(u-1,u)}(1+\theta_{\text{AoI}})=\frac{\lambda_{s}}{\lambda_{s}-\theta_{\text{AoI}}}.
	\end{equation}
	Using Eqs.~(\ref{equation17}) and~(\ref{equation33}), we obtain
	\begin{equation}\label{equation35}
		p_{s}^{\text{AoI}}(u)\leq\frac{\lambda_{s} e^{-\frac{\theta_{\text{AoI}} A_{\text{th}}}{n}}}{\lambda_{s}-\theta_{\text{AoI}}}{\cal M}_{{\cal T}_{s}^{\text{S}}(u)}(1+\theta_{\text{AoI}}), \quad  \forall u.
	\end{equation}

\subsection{The Peak AoI Violation Probability Using  HARQ-IR}

Define $L_{s}$ as the number of successful HARQ-IR retransmissions with $1\leq L_{s}\leq L$.
To calculate the expected value of the average number of successful HARQ-IR retransmissions, denoted as $\mathbb{E}\left[L_{s}\right]$, we derive the probability that the number of HARQ-IR retransmission rounds, when $L_{s}=l$, as follows:
\begin{align}\label{equation0549}
	&\text{Pr}\left\{L_{s}\!=l\right\}\!=\!
	\begin{cases}
		\vspace{3pt}
		\text{Pr}\left\{\overline{{\cal A}}_{0}\right\}\!-\!\text{Pr}\left\{\overline{{\cal A}}_{1}\right\},\qquad\qquad\quad\! \text{for $l=1$}; \\
		\text{Pr}\!\left\{\bigcap\limits_{\iota=1}^{l-1}\left\{\overline{{\cal A}}_{\iota}\right\}\!\!\right\}\!-\!\text{Pr}\!\left\{\bigcap\limits_{\iota=1}^{l}\left\{\overline{{\cal A}}_{\iota}\right\}\!\!\right\}, \text{for $1\!<\!l\!<\!L$}; \\
		\text{Pr}\left\{\bigcap\limits_{l=1}^{L-1}\left\{\overline{{\cal A}}_{l}\right\}\right\}, \qquad\qquad\qquad \text{for $l=L$},
	\end{cases}
\end{align}
where $\overline{{\cal A}}_{l}$ represents the event in which the received packet cannot be successfully decoded after HARQ-IR retransmission $l$.
Accordingly, defining $\epsilon_{s,l}(u)$ as the decoding error probability after HARQ-IR retransmission round $l$, we can obtain the upper bound for the average number of HARQ-IR retransmissions as follows:
\begin{align}\label{equation31}
	\mathbb{E}[L_{s}]&=\sum_{l=1}^{L}l\text{Pr}\left\{L_{s}=l\right\}=\text{Pr}\left\{\,\overline{{\cal A}}_{0}\right\}+\sum_{l=1}^{L-1}\text{Pr}\left\{\!\bigcap\limits_{l=1}^{L-1}\!\left\{\overline{{\cal A}}_{l}\right\}\!\right\}
	\nonumber\\
	&\leq 1+\sum_{l=1}^{L-1}\text{Pr}\left\{\,\overline{{\cal A}}_{l}\right\}\approx
	1+\sum_{l=1}^{L-1}\epsilon_{s,l}(u)
\end{align}
where $\text{Pr}\left\{\,\overline{{\cal A}}_{0}\right\}=1$ and $\epsilon_{s,l}(u)$ denotes the decoding error probability after HARQ-IR retransmission round $l$.

Accordingly, we can obtain the closed-form expression of the decoding error probability function and the peak AoI violation probability considering our proposed performance modeling schemes as shown in the following theorem.

\begin{theorem}\label{theorem03}
	\textbf{If} the channel code defined by Definition 1 is applied to the proposed performance modeling schemes, \textbf{then} the following claims hold for our developed performance modeling schemes.
	
	\underline{Claim 1.} The decoding error probability $\epsilon_{s,l}(u)$ for our proposed modeling schemes is specified as follows:
	\begin{align}\label{theorem03_eq1}
		\epsilon_{s,l}(u)\!\approx&\, \frac{\alpha_{s}(u)}{\Gamma(k_{I_{\text{a}}})(\eta_{I_{\text{a}}})^{k_{I_{\text{a}}}}} \!\sum_{i=1}^{\infty}\!\frac{(m_{s}(u))_{i}[\delta_{s}(u)]^{i}}{(i!)^{2}[\beta_{s}(u)]^{i+1}}\!\!
		\left[\frac{\beta_{s}(u)\zeta_{\text{low},l}}{\phi_{s}(u){\cal P}_{s}(u)}\!\right]^{i+1}
			\nonumber\\
		&\!\times\!\Gamma(i+1)\sum_{j=0}^{\infty}\frac{1}{\Gamma\left(i+j+2\right)}\left[\frac{\beta_{s}(u)\zeta_{\text{low},l}}{\phi_{s}(u){\cal P}_{s}(u)}\right]^{j}
		\nonumber\\
		&\!\times\!
		\left[\frac{\beta_{s}(u)\zeta_{\text{low},l}}{\phi_{s}(u){\cal P}_{s}(u)}+\frac{1}{\eta_{I_{\text{a}}}}\right]^{-(i+j+k_{I_{\text{a}}}+1)}\!\!\!\!\!\!\!\!\Gamma\left(i+j+k_{I_{\text{a}}}+1\right)
		\nonumber\\
		&\!+\!\Bigg[\frac{1}{2}+\vartheta_{s,l}\sqrt{n}\left(e^{R^{*}_{s,l}}-1\right)\Bigg]
		\frac{\alpha_{s}(u)}{\Gamma(k_{I_{\text{a}}})(\eta_{I_{\text{a}}})^{k_{I_{\text{a}}}}}
		\nonumber\\
		&\!\times\! \sum_{i=1}^{\infty}\frac{(m_{s}(u))_{i}[\delta_{s}(u)]^{i}}{(i!)^{2}[\beta_{s}(u)]^{i+1}}\Gamma(i+1)	\left[\frac{\beta_{s}(u)}{\phi_{s}(u){\cal P}_{s}(u)}\right]^{i+1}
		\nonumber\\
		&\!\times\! \Bigg\{ \zeta_{\text{up},l}^{i+1}\!\sum_{j=0}^{\infty}\frac{1}{\Gamma\left(i\!+\!j\!+\!2\right)}\!\left[\frac{\beta_{s}(u)\zeta_{\text{up},l}}{\phi_{s}(u){\cal P}_{s}(u)}\right]^{j}\!
		\nonumber\\
	&\!\times\!	\left[\frac{\beta_{s}(u)\zeta_{\text{up},l}}{\phi_{s}(u){\cal P}_{s}(u)}+\frac{1}{\eta_{I_{\text{a}}}}\right]^{-(i+j+k_{I_{\text{a}}}+1)}\!\!\Gamma\left(i\!+\!j\!+\!k_{I_{\text{a}}}\!+\!1\right)
		\nonumber\\
		&
		\!-\!\zeta_{\text{low},l}^{i+1}\!\sum_{j=0}^{\infty}\frac{1}{\Gamma\left(i\!+\!j\!+\!2\right)}\left[\frac{\beta_{s}(u)\zeta_{\text{low},l}}{\phi_{s}(u){\cal P}_{s}(u)}\right]^{j}\!
		\Bigg[\frac{\beta_{s}(u)\zeta_{\text{low},l}}{\phi_{s}(u){\cal P}_{s}(u)}
		\nonumber\\
		&
		+\frac{1}{\eta_{I_{\text{a}}}}\Bigg]^{-(i+j+k_{I_{\text{a}}}+1)}\!\!\Gamma\left(i\!+\!j\!+\!k_{I_{\text{a}}}\!+\!1\right)\!\!\Bigg\}\!-\!\vartheta_{s,l}\sqrt{n}\Lambda_{l}(u)
	\end{align}
	where $\Gamma(\cdot)$ is the Gamma function, $k_{I_{\text{a}}}=\frac{\left(\mathbb{E}\left[I_{\text{a}}\right]\right)^{2}}{I\left[(I_{\text{a}})^{2}\right]}$,   $\eta_{I_{\text{a}}}=\frac{I\left[(I_{\text{a}})^{2}\right]}{\mathbb{E}\left[I_{\text{a}}\right]}$, where the mean and variance of $I_{\text{a}}$ are given, respectively, as follows~\cite{6308772}:
	\begin{equation}\label{equation031a}
		\begin{cases}
			\!\mathbb{E}\left[I_{\text{a}}\right]\!=\!2\pi \lambda {\cal P}_{t}(u) \sqrt{\frac{k_{\text{pd}+1}}{2k_{\text{pg}}}}\frac{R_{\text{out}}^{2-\alpha}-R_{\text{in}}^{2-\alpha}}{2-\alpha};\\
			\!\text{Var}\left[I_{\text{a}}\right]\!=\!\pi\lambda [{\cal P}_{t}(u)]^{2}k_{\text{pg}}(1+k_{\text{pg}})\eta_{\text{pg}}^{2}\frac{R_{\text{out}}^{2-2\alpha}-R_{\text{in}}^{2-2\alpha}}{1-\alpha},
		\end{cases}
	\end{equation}
and
\begin{equation}
	\begin{cases}
		\vartheta_{s,l}\triangleq\frac{1}{2\pi\sqrt{2^{2R^{*}_{s,l}-1}}};\\
		\zeta_{\text{low},l}\triangleq2^{R^{*}_{s,l}-1} - \frac{1}{2\vartheta_{s,l}\sqrt{\widehat{n}}};\\
		\zeta_{\text{up},l}\triangleq2^{R^{*}_{s,l}-1} + \frac{1}{2\vartheta_{s,l}\sqrt{\widehat{n}}},
	\end{cases}
\end{equation}
and
\begin{align}\label{equation073b}
	&\Lambda_{l}(u)=
	\frac{\alpha_{s}(u)}{\Gamma(k_{I_{\text{a}}})(\eta_{I_{\text{a}}})^{k_{I_{\text{a}}}}}
	\sum_{j=0}^{m_{s}(u)-1}\frac{(-1)^{j}\left(1-m_{s}(u)\right)_{j}}{(j!)^{2}}
	\nonumber\\
	&\quad\!\!\!\!\!\!\times\!\!\left[ \frac{\delta_{s}(u)}{\phi_{s}(u){\cal P}_{s}(u)}\right]^{j}(j+k_{I_{\text{a}}})!\frac{\eta_{I_{\text{a}}}^{j+k_{I_{\text{a}}}+1}}{j+2} \Bigg[
	\left(\zeta_{\text{up},l}\right)^{j+2}
	\nonumber\\
	&\quad \!\!\!\!\!\!\times\!\!\! \sideset{_2}{_{1}}{\mathop{F}}\!\!\left(j\!+\!k_{I_{\text{a}}}\!+\!1,\!j\!+\!2,\!j\!+\!3,\!-\frac{\zeta_{\text{up},l}\eta_{I_{\text{a}}}\left[\beta_{s}\!(u)\!-\!\delta_{s}\!(u)\right]}{\phi_{s}(u){\cal P}_{s}(u)}\right)
	\nonumber\\
	&\quad\!\!\!\!\!\!- \!\! \sideset{_2}{_{1}}{\mathop{F}}\!\!\left(\!j\!+\!k_{I_{\text{a}}}\!+\!1,j\!+\!2,j\!+\!3,-\frac{\eta_{I_{\text{a}}}\left[\beta_{s}(u)-\delta_{s}(u)\right]}{\phi_{s}(u){\cal P}_{s}(u)}\zeta_{\text{low},l}\!\right)
\nonumber\\
&\quad\!\!\!\!\!\!\times  \!\left(\zeta_{\text{low},l}\right)^{j+2}\Bigg]
\end{align}
{where $_{2}F_{1}(g,b,s,z)$ is the hypergeometric function, which is defined as follows:
\begin{equation}
\sideset{_2}{_{1}}{\mathop{F}}(g,b,s,z)=\sum_{\ell=0}^{\infty}\frac{(g)_{\ell}(b)_{\ell}}{(s)_{\ell}}\frac{z^{\ell}}{\ell!}
	\end{equation}
	where $(q)_\ell$ $(q\in\{g,b,s\})$ is the Pochhammer symbol, which is defined by:
	\begin{equation}
		(q)_\ell=
		\begin{cases}
			1, \qquad\qquad\qquad\qquad\qquad \ell = 0;\\
			q(q+1)\cdots (q+\ell-1),\quad \ell >0.
		\end{cases}
	\end{equation}
}

	\underline{Claim 2.} The peak AoI violation probability $p_{s}^{\text{AoI}}(u)$ is derived by using the decoding error probability $\epsilon_{s,l}(u)$ as follows:
	\begin{equation}\label{theorem03_eq2a}
		p_{s,l}^{\text{AoI}}(u)\approx\frac{\lambda_{s} e^{-\frac{\theta_{\text{AoI}} A_{\text{th}}}{\widehat{n}}}}{\lambda_{s}-\theta_{\text{AoI}}}\exp\left\{ \theta_{\text{AoI}}\widehat{n}T\left[	1\!+\!\sum_{l=1}^{L-1}\epsilon_{s,l}(u)\right]\right\}
	\end{equation}
	where $T$ is the unit time for each channel use.
	The peak AoI violation probability $p_{s}^{\text{AoI}}(u)$ by implementing HARQ-IR for the developed modeling formulations/frameworks is specified as follows:
\begin{align}\label{theorem03_eq2}
	p_{s,l}^{\text{AoI}}(u)\approx&\frac{\lambda_{s} e^{-\frac{\theta_{\text{AoI}} A_{\text{th}}}{\widehat{n}}}}{\lambda_{s}-\theta_{\text{AoI}}}\exp\left\{ \theta_{\text{AoI}}\widehat{n}T\left[	1\!+\!\sum_{l=1}^{L-1}\epsilon_{s,l}(u)\right]\right\}
	\nonumber\\
	\approx&\frac{\lambda_{s} e^{-\frac{\theta_{\text{AoI}} A_{\text{th}}}{\widehat{n}}}}{\lambda_{s}-\theta_{\text{AoI}}}\exp\Bigg\{ \theta_{\text{AoI}}\widehat{n}T\Bigg[	1+\sum_{l=1}^{L-1} \frac{\alpha_{s}(u)}{\Gamma(k_{I_{\text{a}}})(\eta_{I_{\text{a}}})^{k_{I_{\text{a}}}}} 	\nonumber\\
	&\!\!\times\!\sum_{i=1}^{\infty}\frac{(m_{s}(u))_{i}[\delta_{s}(u)]^{i}}{(i!)^{2}[\beta_{s}(u)]^{i+1}}
	\left[\frac{\beta_{s}(u)\zeta_{\text{low},l}}{\phi_{s}(u){\cal P}_{s}(u)}\right]^{i+1}
	\nonumber\\
	&\!\!\times\! \Gamma(i\!+\!1)\!\sum_{j=0}^{\infty}\!\frac{1}{\Gamma\left(i\!+\!j\!+\!2\right)}\left[\frac{\beta_{s}(u)\zeta_{\text{low},l}}{\phi_{s}(u){\cal P}_{s}(u)}\right]^{j}\!
	\nonumber\\
	&\!\!\times\!\left[\frac{\beta_{s}(u)\zeta_{\text{low},l}}{\phi_{s}(u){\cal P}_{s}(u)}+\frac{1}{\eta_{I_{\text{a}}}}\right]^{-(i+j+k_{I_{\text{a}}}+1)}\!\!\!\!\!\!\!\!\!\Gamma\left(i\!+\!j\!+\!k_{I_{\text{a}}}\!+\!1\right)
	\nonumber\\
	&\!\!+\!\Bigg[\frac{1}{2}+\vartheta_{s,l}\sqrt{\widehat{n}}\left(e^{R^{*}_{s,l}}-1\right)\Bigg]
	\frac{\alpha_{s}(u)}{\Gamma(k_{I_{\text{a}}})(\eta_{I_{\text{a}}})^{k_{I_{\text{a}}}}}
	\nonumber\\
	&\!\!\times\!\sum_{i=1}^{\infty}\frac{(m_{s}(u))_{i}[\delta_{s}(u)]^{i}}{(i!)^{2}[\beta_{s}(u)]^{i+1}}\Gamma(i\!+\!1)\!	\left[\frac{\beta_{s}(u)}{\phi_{s}(u){\cal P}_{s}(u)}\right]^{i+1}
	\nonumber\\
	&\!\!\times\! \Bigg\{ \zeta_{\text{up},l}^{i+1}\!\sum_{j=0}^{\infty}\!\frac{1}{\Gamma\left(i\!+\!j\!+\!2\right)}\left[\frac{\beta_{s}(u)\zeta_{\text{up},l}}{\phi_{s}(u){\cal P}_{s}(u)}\right]^{j}
	\nonumber\\
	&\!\!\times\!
	\left[\frac{\beta_{s}(u)\zeta_{\text{up},l}}{\phi_{s}(u){\cal P}_{s}(u)}+\frac{1}{\eta_{I_{\text{a}}}}\right]^{-(i+j+k_{I_{\text{a}}}+1)}\!\!\!\!\!\!\!\!\!\Gamma\left(i\!+\!j\!+\!k_{I_{\text{a}}}\!+\!1\right)
	\nonumber\\
	&
	\!\!-\!\zeta_{\text{low},l}^{i+1}\!\sum_{j=0}^{\infty}\!\frac{1}{\Gamma\!\left(i\!+\!j\!+\!2\right)}\!\!\left[\frac{\beta_{s}(u)\zeta_{\text{low},l}}{\phi_{s}(u){\cal P}_{s}(u)}\right]^{\!j}\!\!
	\Bigg[\!\frac{\beta_{s}(u)\zeta_{\text{low},l}}{\phi_{s}(u){\cal P}_{s}(u)}
	\nonumber\\
	&\!\!+\!\frac{1}{\eta_{I_{\text{a}}}}\!\Bigg]^{-(i+j+k_{I_{\text{a}}}+1)}\!\!\!\!\!\!\!\!\!\!\Gamma\left(i\!+\!j\!+\!k_{I_{\text{a}}}\!+\!1\right)\!\!\!\Bigg\}\!
	-\!\vartheta_{s,l}\sqrt{\widehat{n}}\Lambda_{l}(u)\!\Bigg]\!\Bigg\}.
\end{align}
\end{theorem}

\begin{IEEEproof}
				The proof of Theorem~\ref{theorem03} is in Appendix C.
\end{IEEEproof}

\indent\textit{Remarks on Theorem~\ref{theorem03}:} Theorem~\ref{theorem03} provides an analysis of peak AoI violation probability by determining the decoding error probability.
Due to the complexity of the obtained closed-form expression for the decoding error probability, we opt for a more convenient approach by leveraging its asymptotic representation in the high SNR regime in the following, simplifying the analysis to achieve a clearer understanding of the system's behavior while retaining its essential characteristics.

\subsection{Asymptotic Peak AoI Violation Probability}
Considering high SNR region, i.e., ${\cal P}_{s}(u)\rightarrow\infty$, we can derive the asymptotic cumulative probability function (CDF) of the channel gain $|h_{s}(u)|^{2}$, denoted by $F^{\infty}_{|h_{s}(u)|^{2}}(x)$, between the satellite and the destination node as follows:
\begin{equation}\label{equation074}
	F^{\infty}_{|h_{s}(u)|^{2}}\!(x)\!=\!\alpha_{s}(u)\! \!\sum_{i=1}^{\infty}\!\frac{(m_{s}(u)\!)_{i}[\delta_{s}(u)]^{i}}{(i!)^{2}[\beta_{s}(u)]^{i+1}}\frac{\left[\beta_{s}(u) x\right]^{i+1}}{i+1}
	\!\approx\! \alpha_{s}(u)x.
\end{equation}
Given the distribution of the interference power $I_{\text{a}}$, the asymptotic CDF of the SINR, denoted by $F^{\infty}_{\gamma_{s}(u)}(x)$, between the satellite and the destination node can be derived as follows:
\begin{align}\label{equation075}
	F^{\infty}_{\gamma_{s}(u)}(x)
	&=\int_{0}^{\infty}F^{\infty}_{|h_{s}(u)|^{2}}\left(\frac{xy}{\phi_{s}(u){\cal P}_{s}(u)}\right)f_{I_{\text{a}}}(y)dy
	\nonumber\\
	&=\frac{\eta_{I_{\text{a}}}^{k_{I_{\text{a}}}+1}\Gamma\left(k_{I_{\text{a}}}+1\right)\alpha_{s}(u)x}
	{\Gamma(k_{I_{\text{a}}})(\eta_{I_{\text{a}}})^{k_{I_{\text{a}}}}\phi_{s}(u){\cal P}_{s}(u)}.
\end{align}
In the high SNR region, we obtain the asymptotic peak AoI violation probability, as elucidated in the following lemma.

\textit{Lemma 2:}
	The asymptotic peak AoI violation probability, denoted by $p_{s,l}^{\text{AoI},\infty}(u)$, after HARQ-IR retransmission round $l$ in the high SNR region is given as follows:
	\begin{align}\label{lemma02_eq01}
	p_{s,l}^{\text{AoI},\infty}\!(u)\!\approx&\,\frac{\lambda_{s} e^{-\frac{\theta_{\text{AoI}} A_{\text{th}}}{\widehat{n}}}}{\lambda_{s}-\theta_{\text{AoI}}}\exp\!\Bigg\{\!\theta_{\text{AoI}}\widehat{n}T\Bigg[\!1\!+\!
	\!\sum_{l=1}^{L-1}\frac{\eta_{I_{\text{a}}}^{k_{I_{\text{a}}}+1}\Gamma\!\left(k_{I_{\text{a}}}\!+\!1\right)}
	{\Gamma(k_{I_{\text{a}}})(\eta_{I_{\text{a}}})^{k_{I_{\text{a}}}}}
	\nonumber\\
	&\!\times\!\frac{\alpha_{s}(u)}
	{\phi_{s}(u){\cal P}_{s}(u)}
	\Bigg\{\zeta_{\text{low},l}\!+\!\Bigg[\frac{1}{2}+\!\vartheta_{s,l}\sqrt{\widehat{n}}\!\left(e^{R^{*}_{s,l}}\!-\!1\right)\!\!\Bigg]
	\nonumber\\
	&\!\!\times \!\left(\zeta_{\text{up},l}\!-\!\zeta_{\text{low},l}\right)
	\!-\!\vartheta_{s,l}\sqrt{\widehat{n}}\left(\zeta_{\text{up},l}^{2}\!-\!\zeta_{\text{low},l}^{2}\right)^{2}\!
		\nonumber\\
	&\!\!\times\!
	\!\left[1\!-\!\frac{\eta_{I_{\text{a}}}^{k_{I_{\text{a}}}+1}\Gamma\left(k_{I_{\text{a}}}\!+\!1\right)\alpha_{s}(u)}
	{2\Gamma(k_{I_{\text{a}}})(\eta_{I_{\text{a}}})^{k_{I_{\text{a}}}}\phi_{s}(u){\cal P}_{s}(u)}\!\right]\!
	\!\Bigg\}\!\Bigg]\!\Bigg\}.
\end{align}

\begin{IEEEproof}
				The proof of Lemma~2 is in Appendix D.
\end{IEEEproof}

\indent\textit{Remarks on Lemma~2:} Lemma~2 provides an analysis of the asymptotic peak AoI violation probability by determining the asymptotic decoding error probability for our developed modeling schemes employing HARQ-IR.
We can observe that the asymptotic peak AoI violation probability in the high SNR region is a function in terms of the blocklength, the peak-AoI bounded QoS exponent, and the rate of the arrival updates.
The asymptotic peak-AoI violation probability is a monotonically increasing function of blocklength $\widehat{n}$. 

\section{Statistical Delay and Error-Rate Bounded QoS Metrics Over Satellite-Terrestrial Integrated Networks Using FBC}\label{sec:EC1}

In this section,  we develop a set of new fundamental statistical QoS controlling metrics for the developed satellite-terrestrial integrated schemes using FBC.
\subsection{The Statistical Delay-Bounded QoS Metric}
Research has been conducted on statistical delay-bounded QoS guarantees~\cite{C1994,chang2000performance} to investigate queuing dynamics over arrival and service processes that vary over time.

\textit{Definition 3. The statistical delay-bounded QoS exponent:} By using the LDP, the queueing process converges to a random variable $Q_{s}(\infty)$ in distribution for satellite-terrestrial integrated wireless communications such that
\begin{equation}\label{equation23}
	-\lim_{Q_{\text{th},s}\rightarrow\infty}\frac{\log\left(\text{Pr}
		\left\{Q_{s}(\infty)>Q_{\text{th},s}\right\}\right)}{Q_{\text{th},s}}=\theta_{\text{delay}}
\end{equation}
where $Q_{\text{th},s}$ denotes the threshold of buffer overflow and $\theta_{\text{delay}}$ $(\theta_{\text{delay}}>0)$ represents the delay-bounded QoS exponent to describe the queuing delay, where it quantifies the rate of exponential decay that occurs in the delay-bounded QoS violation probabilities. 		
Specifically, a sufficiently small delay-bounded QoS exponent enables the satellite-terrestrial integrated system to accommodate an arbitrarily prolonged delay, whereas a sufficiently large delay-bounded QoS exponent renders the system intolerant of any delay.

In addition, we define $D_{s}(u)$ is the random delay process.
Denote by $D_{\text{th},s}$ the delay constraint.
By leveraging the principles of statistical QoS theory, we can demonstrate the following approximate derivation of the probability of violating the upper bound for queueing delay, denoted by $\text{Pr}\left\{D_{s}(u)\geq D_{\text{th},s}\right\}$:
\begin{align}\label{equation062}
	\text{Pr}\left\{D_{s}(u)\geq D_{\text{th},s}\right\}&\approx
	\Delta_{s} \exp\left\{-\theta_{\text{delay}}R_{s}(u)D_{\text{th},s}\right\}
\end{align}
where $\Delta_{s}\triangleq \text{Pr}\left\{Q_{s}(u)>0\right\}$ is the probability of a non-empty queue and $R_{s}(u)$  $(l=1,2,\dots)$ is service process.


Furthermore, considering the bit domain, we can derive the service process, denoted by $S_{s}(u)$, during the transmission of status update $u$ as follows:			
	\begin{equation}\label{equation082}
		S_{s}(u)=
		\begin{cases}
			\log_{2}(M^{*}_{s}), \qquad\text{with probability } 1-\epsilon_{s}(u);\\
			0,  \qquad\qquad\qquad\!\! \text{with probability } \epsilon_{s}(u).
		\end{cases}
	\end{equation}
Correspondingly, the Mellin transform in terms of the service process, denoted as ${\cal M}_{{\cal S}_{s}(u)}(\theta_{\text{delay}})$, is derived as follows:
	\begin{align}\label{equation83}
		{\cal M}_{{\cal S}_{s}(u)}(\theta_{\text{delay}})
		&=\mathbb{E}_{\gamma_{s}(u)}\left[\epsilon\left(\gamma_{s}(u)\right)\right] +\mathbb{E}_{\gamma_{s}(u)}\left[1-\epsilon\left(\!\gamma_{s}(u)\right)\right]
		\nonumber\\
		&\quad\times e^{(\theta_{\text{delay}}-1)  \log_{2}(M^{*}_{s})}.
	\end{align}
Define ${\cal M}_{{\cal A}_{s}}(\theta_{\text{delay}})$ as the Mellin transform in terms of the arrival process $A_{s}(u)$ in the SNR domain.
Under the assumption of i.i.d. arrivals, which implies that the accumulated source rate $A_{s}(u)$ exhibits i.i.d. increments, represented as $a_{s}(u)$ or equivalently, denoted by $a_{s}=a_{s}(u)$, owing to the i.i.d. nature of $a_{s}(u)$.
Accordingly, we can proceed to determine the Mellin transform of the accumulated arrival process, denoted as ${\cal M}_{{\cal A}_{s}}(\theta_{\text{delay}})$, as outlined below:
	\begin{align}\label{equation031}
		{\cal M}_{{\cal A}_{s}}(\theta_{\text{delay}})&\!=\!\mathbb{E}\!\left[\!\left(\prod_{j=1}^{u}e^{a_{s}(j)}\right)^{\theta_{\text{delay}}-1}\right]
		\!=\!\left(\mathbb{E}\!\left[e^{a_{s}\left(\theta_{\text{delay}}-1\right)}\right]\right)^{\!\!u}
		\nonumber\\
		&\!=\!\left[{\cal M}_{\widetilde{a}_{s}}(\theta_{\text{delay}})\right]^{u}
	\end{align}	
	where $\widetilde{a}_{s}=e^{a_{s}}$.
Consider the scenario where the arrival is modeled as a Poisson distribution with a rate parameter $\lambda_{s}$.
Under this assumption, we can proceed to obtain the Mellin transform of $\widetilde{a}_{s}$ through the following derivation:
	\begin{align}\label{equation24}
		{\cal M}_{\widetilde{a}_{s}}(\theta_{\text{delay}})=\sum_{i=1}^{\infty}e^{i(\theta_{\text{delay}}-1)}\frac{\left(\lambda_{s}\right)^{i}}{i!}e^{-\lambda_{s}}=e^{\lambda_{s}\left(e^{\theta_{\text{delay}}-1}-1\right)}.
	\end{align}	
Given the target delay $D_{\text{th}}$, a kernel function $\widetilde{\mathsf{K}}_{s}(\theta_{\text{delay}},u)$ for measuring the queuing delay is defined as follows~\cite{HAL2016}:
	\begin{equation}\label{equation1126}
		\widetilde{{\mathsf{K}}}^{(u)}(\theta_{\text{delay}},D_{\text{th},s})\!\triangleq\! \frac{\left[{\cal M}_{{\cal S}_{s}(u)}(1-\theta_{\text{delay}})\right]^{D_{\text{th},s}}}
		{1-{\cal M}_{{\cal A}_{s}}(1+\theta_{\text{delay}}){\cal M}_{{\cal S}_{s}(u)}(1-\theta_{\text{delay}})}
	\end{equation}
if the following stability condition can hold:
	\begin{equation}
		{\cal M}_{{\cal A}_{s}}(\theta_{\text{delay}})(1+\theta_{\text{delay}}){\cal M}_{{\cal S}_{s}(u)}(1-\theta_{\text{delay}})<1.
	\end{equation}
By employing the Mellin transform in the SNR domain, the arrival process $A_{s}(u)$ and service process $S_{s}(u)$ can be effectively analyzed. Thus, the upper-bounded delay violation probability, denoted by $p_{s}^{\text{delay}}(u)$, is given as follows:
	\begin{equation}\label{equation127}
		p_{s}^{\text{delay}}(u)\leq  \inf_{\theta_{\text{delay}}>0}\left\{\widetilde{\mathsf{K}}^{(u)}(\theta_{\text{delay}}, D_{\text{th},s})\right\}.
	\end{equation}

	\subsection{The Error-Rate Bounded QoS Metric}
We investigate the statistical QoS requirement that describes the tail behavior of the decoding error probability with an increasing blocklength.
In the realm of small-packet communications, the tradeoff between reliability and throughput has conventionally been studied by characterizing the \textit{error-rate bounded QoS exponent}.
Consequently, 
to represent error-rate bounded QoS metrics over satellite-terrestrial integrated wireless networks, it is crucial to explore the connections between the \textit{error-rate bounded QoS exponent} and the maximum achievable coding rate.

\textit{Definition 4: The error-rate bounded QoS exponent:} Based on the LDP, the \textit{error-rate bounded QoS exponent}, denoted by $\theta_{\text{error}}$, quantifies the rate of exponential decay of the QoS violation probability with respect to the decoding error probability $\epsilon(\mu)$, within the framework of our proposed modeling schemes, which is defined as follows~\cite{gallager1968information}:
\begin{equation}\label{equation38}
	\theta_{\text{error}}\triangleq\lim\limits_{n\rightarrow \infty}-\frac{1}{n}\log(\epsilon_{s}(u)).
\end{equation}
Our proposal centers on the modeling and assessment of fundamental performance, with a particular emphasis on various facets related to the error-rate bounded QoS exponent. This exponent delineates the rate at which the decoding error probability exponentially decays as the codeword blocklength tends to infinity, i.e.,
\begin{equation}\label{equation039}
	\epsilon_{s}(u)\leq\exp(-n\theta_{\text{error}}).
\end{equation}
Based on the definition in~\cite{gallager1968information}, the \textit{error-rate bounded QoS exponent} $\theta_{\text{error}}$ is obtained as follows:
\begin{equation}\label{equation040}
\theta_{\text{error}}\triangleq\sup_{ \rho_{s}\in[0,1]}\left\{E_{0}\left[\rho_{s},P_{\bm{x}_{s}}\left(\bm{x}_{s}\right)\right]-\rho_{s} R^{*}_{s}\right\}
\end{equation}
where $\rho_{s}\in[0,1]$ represents the Lagrange multiplier parameter, $P_{\bm{x}_{s}}\left(\bm{x}_{s}\right)$ is the PDF of the transmitted signal vector $\bm{x}_{s}$, and
\begin{align}\label{equation041}
	&E_{0}\left[\rho_{s},P_{\bm{x}_{s}}\left(\bm{x}_{s}\right)\right]\triangleq -\frac{1}{n}\log\! \Bigg\{\!\mathbb{E}_{\gamma_{s}(u)}\!\Bigg[\int_{\bm{y}_{s}}\!\Bigg[\int_{\bm{x}_{s}}\!\!P_{\bm{x}_{s}}\!\left(\bm{x}_{s}\right)
	\nonumber\\
	&\quad\times \left[P_{\bm{y}_{s}|\bm{x}_{s},h_{s}}\!\left(\bm{y}_{s}|\bm{x}_{s},h_{s}\right)\right]^{\frac{1}{(1+\rho_{s})}}d\bm{x}_{s}\Bigg]^{(1+\rho_{s})}\!\!\!d\bm{y}_{s}\!\Bigg]\!\Bigg\}\!.
\end{align}
where $\bm{y}_{s}$ denotes the received signal vector at the destination GBS.
Thus, we need to derive the error-rate bounded QoS exponent defined by Eq.~\eqref{equation040} for modeling the error-rate bounded QoS metrics by implementing FBC techniques.
By substituting the conditional PDF $P_{\mathbf{Y}|\bm{x}_{s},h_{s}}\left(\mathbf{Y}|\bm{x}_{s},h_{s}\right)$ into Eq.~\eqref{equation041}, the following equation can be obtained~\cite{Gallager2009}:
\begin{align}\label{equation047}
	E_{0}\left[\rho_{s},P_{\bm{x}_{s}}\left(\bm{x}_{s}\right)\right]&\!=\!-\frac{1}{n}\log \left\{\mathbb{E}_{\gamma_{s}(u)}\!\left[\left(1\!+\!\frac{\gamma_{s}(u)}{1+\rho_{s}}\right)^{-n\rho_{s}}\right]\right\}.
\end{align}
	Then, plugging Eq.~\eqref{equation047} in Eq.~\eqref{equation040}, the error-rate bounded QoS exponent $\theta_{\text{error}}$ is determined for the proposed performance modeling frameworks as follows:
\begin{align}\label{equation046}
	\theta_{\text{error}}\!=&\!\sup_{ \rho_{s}\in[0,1]}\Bigg\{-\frac{1}{n}\log \left\{\mathbb{E}_{\gamma_{s}(u)}\left[\left(1\!+\!\frac{\gamma_{s}(u)}{1+\rho_{s}}\right)^{-n\rho_{s}}\right]\right\}
	\nonumber\\
	&\qquad\qquad-\rho_{s} R^{*}_{s}\Bigg\}.
\end{align}

However, how to analytically derive the closed-form expression of the error-rate bounded QoS exponent through determining the term $E_{0}\left[\rho_{s},P_{\bm{x}_{s}}\left(\bm{x}_{s}\right)\right]$ in Eq.~\eqref{equation047} over satellite-terrestrial integrated wireless networks remains an open and challenging problem.
Towards this end, by applying Jensen's inequality, we can derive the asymptotic $E_{0}\left[\rho_{s},P_{\bm{x}_{s}}\left(\bm{x}_{s}\right)\right]$ specified by Eq.~\eqref{equation047} between the satellite and the destination node which are summarized in the following corollary.

\textit{Corollary 1:}
	\textbf{If} our developed satellite-terrestrial integrated performance modeling scheme is applied for statistical QoS theory, \textbf{then} the asymptotic $E_{0}\left[\rho_{s},P_{\bm{x}_{s}}\left(\bm{x}_{s}\right)\right]$ is determined as follows:
	\begin{align}\label{theorem04_eq00}
		&E_{0}\left[\rho_{s},P_{\bm{x}_{s}}\left(\bm{x}_{s}\right)\right]
		\nonumber\\
		&\quad\approx\!\rho_{s}\!\log\!\Bigg\{\!\phi_{s}(u){\cal P}_{s}(u)\alpha_{s}(u)\!\!\left[\Gamma(2)\! \sideset{_2}{_{1}}{\mathop{F}}\!\!\left(\!m_{s}(u),2;1;\frac{\delta_{s}(u)}{\beta_{s}(u)}\!\right)\!\right]
		\nonumber\\
		&\qquad\times\! [\beta_{s}(u)]^{2} \! +\!(1\!+\!\rho_{s})\!\sqrt{\frac{\pi}{2}}\varsigma\!\sum\limits_{j=1}^{K}\!\phi_{j}(u)P_{t}(u)+\!1\!+\!\rho_{s}\Bigg\}
		\nonumber\\
		&\qquad-\rho_{s}\psi\left(k_{I_{\text{a}}}\right)-\rho_{s}\log\left(\eta_{I_{\text{a}}}\right)
		-\rho_{s}\log \left(1+\rho_{s}\right)
	\end{align}
where $\psi(\cdot)$ is the Digamma function.

\begin{IEEEproof}
						The proof of Corollary~1 is in Appendix E.
\end{IEEEproof}

\indent\textit{Remarks on Corollary 1}: Corollary 1 studies the term $E_{0}\left[\rho_{s},P_{\bm{x}_{s}}\left(\bm{x}_{s}\right)\right]$, which can then be applied in analyzing the error-rate bounded QoS exponent $\theta_{\text{error}}$.
Moreover, the developed statistical delay and error-rate bounded QoS provisioning schemes can be well-suited for deployment in dense scenarios, particularly in densely populated urban areas. 
This is attributed to their formulation and implementation of QoS standards based on the channel feedback. With an accurate channel model and interference model in place, the fundamental statistical QoS metrics can be established to effectively mitigate tail behaviors associated with delay and reliability.

{Furthermore, to conduct a comprehensive evaluation of the effectiveness of the developed fundamental statistical delay and error-rate bounded QoS controlling metrics, it is important to explore strategies aimed at enhancing system performance through interference mitigation since the interference poses a significant challenge within the satellite-terrestrial integrated networks, impacting signal quality and overall system reliability. 
Therefore, the mitigation of interference emerges as a critical aspect for optimizing system performance and achieving the desired QoS metrics.
In particular,} Fractional Frequency Reuse (FFR)~\cite{6047548} has been proposed to serve as an interference coordination static mechanism employed to mitigate cell-edge interference within networks adhering to a frequency reuse of 1. 
It achieves this by constraining transmissions in distinct segments of the cell-edge areas to specific portions of the spectrum channel utilized in the network. 
In addition, strategies involving dynamic power control~\cite{5949138} entail adjusting the transmission power in accordance with the received signal strength. 
This approach contributes to maintaining a harmonious equilibrium between coverage and interference. 
Additionally, intelligent scheduling algorithms~\cite{caglar2016intelligent} consider interference patterns, strategically optimizing resource allocation to mitigate the effects of interference.
{Through the examination of interference mitigation techniques within the framework of our developed QoS controlling metrics, we can evaluate their effectiveness in reducing interference and enhancing system performance. 
	Consequently, by analyzing the interplay between interference mitigation strategies and the developed QoS metrics, we can then identify optimal configurations and operational parameters that maximize system performance, including throughput and system capacity, all while ensuring QoS requirements.}

	\section{Performance Evaluations}\label{sec:results}
	We present a series of numerical results to verify and assess the effectiveness of our proposed statistical QoS provisioning schemes over satellite-terrestrial integrated 6G mobile wireless networks using FBC.
	Throughout our simulations, we set the outer radius $R_{\text{out}}=10$ Km, inner radius $R_{\text{in}}=2$ Km,  the antenna gain at satellite $G_{s}=20$ dBi, and the transmit power at the satellite $P_{s}(u)\in[10,50]$ dBm.

	\begin{figure}[!t]
		\vspace{-5pt}
		\centering
		\includegraphics[scale=0.34]{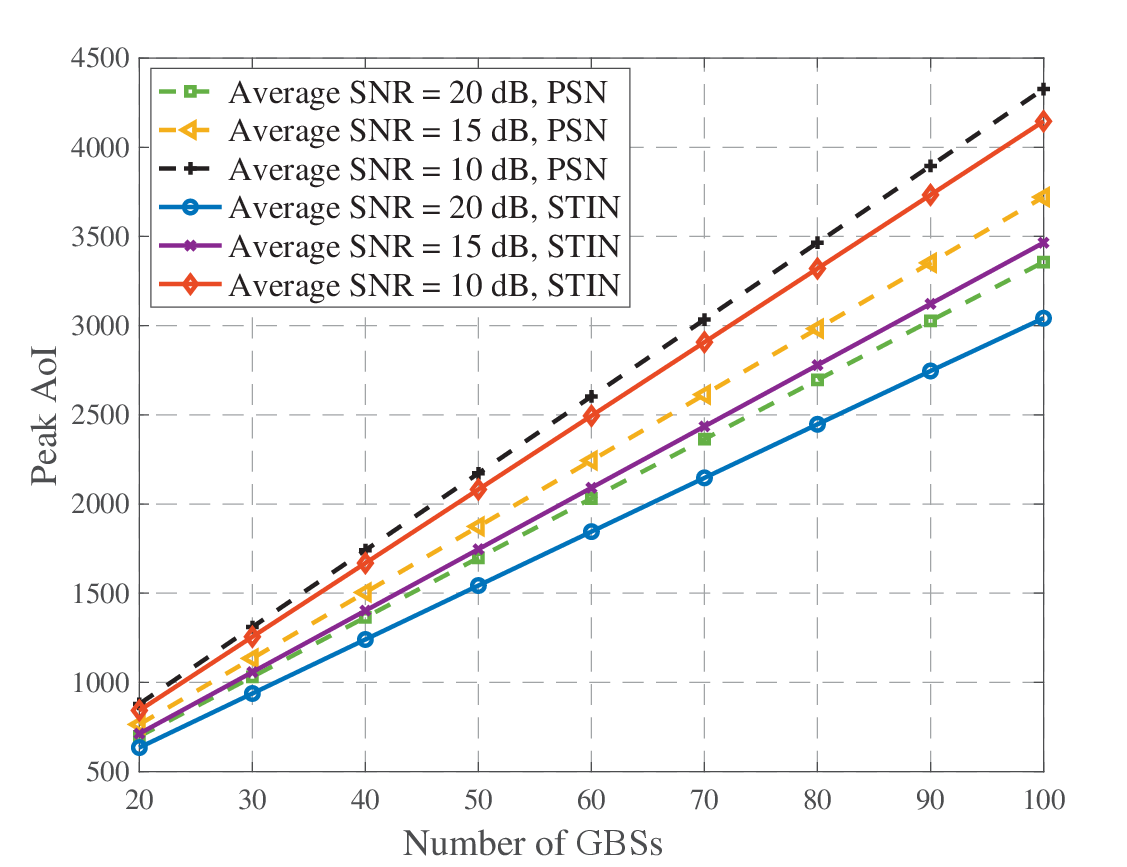}
		\caption{The peak AoI vs. the number of GBSs for our developed satellite-terrestrial integrated networks.}
		\label{fig:03}
		\vspace{-5pt}
	\end{figure}
	
	Figure~\ref{fig:03} depicts the peak AoI against the number of GBSs for the developed satellite-terrestrial integrated networks (STIN) as compared with the pure satellite networks (PSN).
Fig.~\ref{fig:03} illustrates the peak AoI exhibits an increasing trend with respect to the number of GBSs, which is attributed to the fact that a larger number of GBSs results in a corresponding escalation of queuing delays.
	Fig.~\ref{fig:03} also shows that the proposed STIN achieves better peak AoI performance as compared to the PSN since the peak AoI in terrestrial systems is relatively lower than that of the PSN.
	
	\begin{figure}[!t]
		\vspace{-5pt}
		\centering
		\includegraphics[scale=0.34]{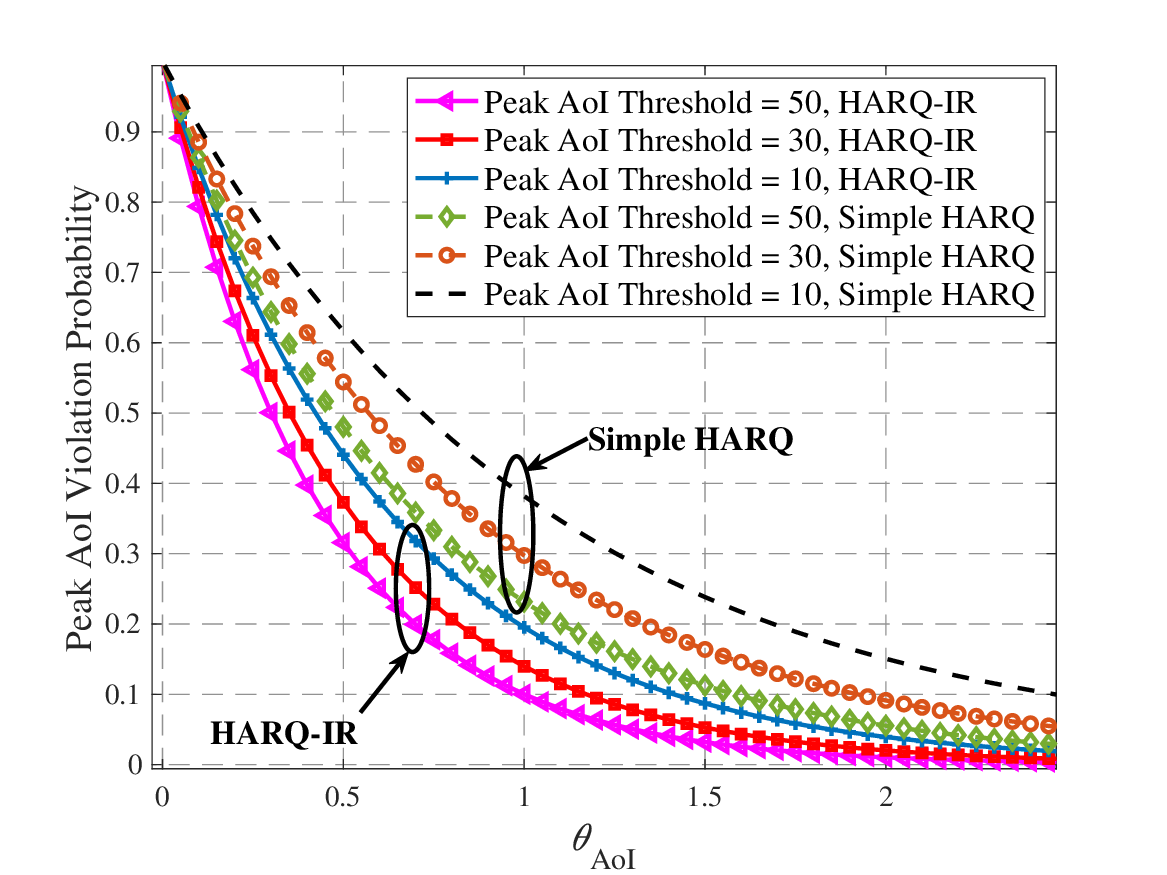}
		\caption{The peak AoI violation probability vs. the peak-AoI bounded QoS exponent $\theta_{\text{AoI}}$ for our developed satellite-terrestrial integrated networks.}
		\label{fig:04}
		\vspace{-5pt}
	\end{figure}

\begin{figure}[!t]
		\vspace{-5pt}
	\centering
	\includegraphics[scale=0.3]{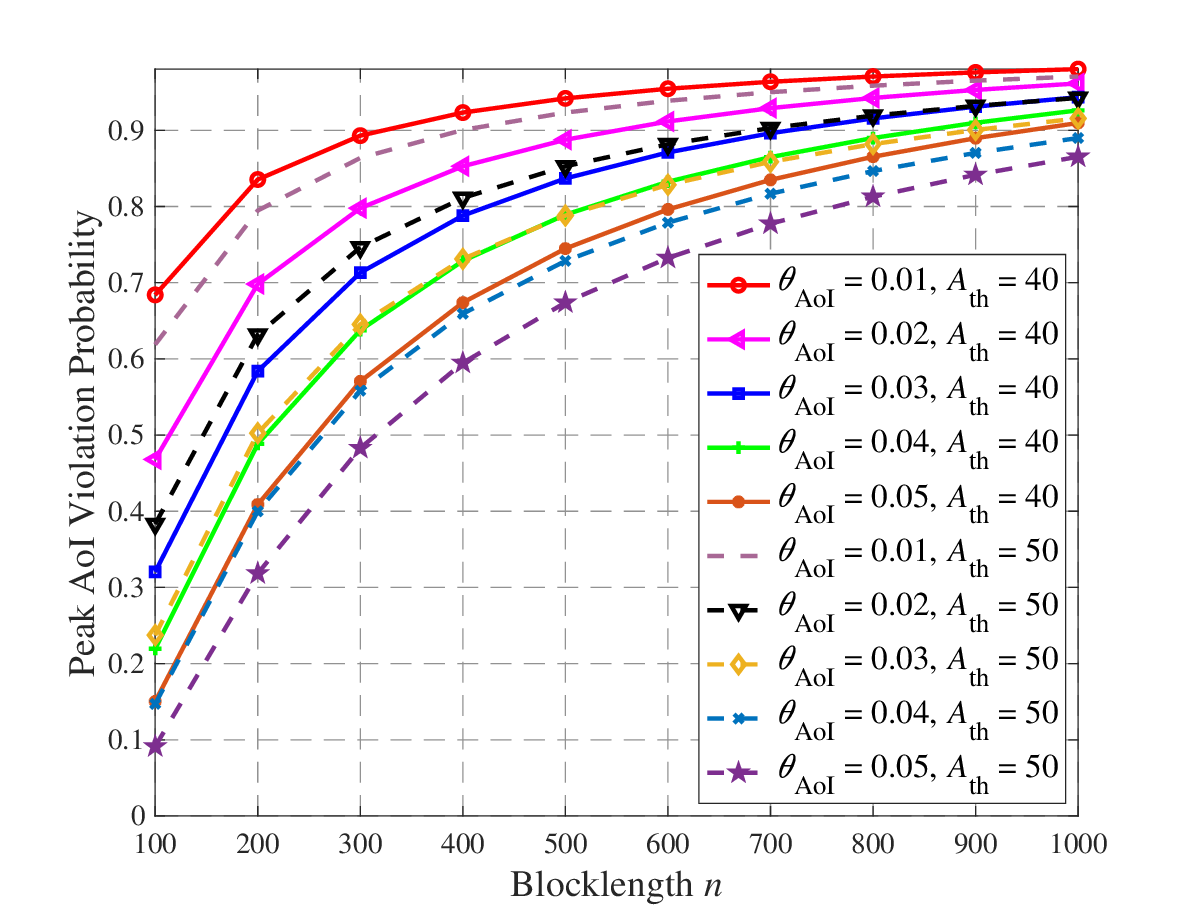}
	\caption{The peak AoI violation probability vs. the blocklength for the developed satellite-terrestrial integrated networks.}
	\label{fig:04b}
	\vspace{-5pt}
\end{figure}

Figure~\ref{fig:04} plots the peak AoI violation probability against the peak-AoI bounded QoS exponent $\theta_{\text{AoI}}$ using HARQ-IR as compared with the simple HARQ scheme.
Fig.~\ref{fig:04} demonstrates a decreasing trend in the peak AoI violation probability when $\theta_{\text{AoI}}$ varies.
This observation serves as confirmation of the reliability and accuracy of $\theta_{\text{AoI}}$.
Notably, a large peak-AoI bounded QoS exponent establishes the lower bound, whereas a small peak-AoI bounded QoS exponent sets the upper bound on the peak AoI violation probability.
Fig.~\ref{fig:04} also shows that the HARQ-IR scheme outperforms the simple HARQ scheme in terms of the peak AoI violation probability.
Moreover, Fig.~\ref{fig:04b} depicts the peak AoI violation probability against blocklength.
Fig.~\ref{fig:04b} indicates that the peak AoI violation probability exhibits an increasing trend against blocklength $n$.
The plot in Fig.~\ref{fig:04b} demonstrates that a larger AoI threshold $A_{\text{th},s}$ leads to a smaller peak-AoI violation probability.

\begin{figure}[!t]
		\vspace{-5pt}
		\centering
		\includegraphics[scale=0.31]{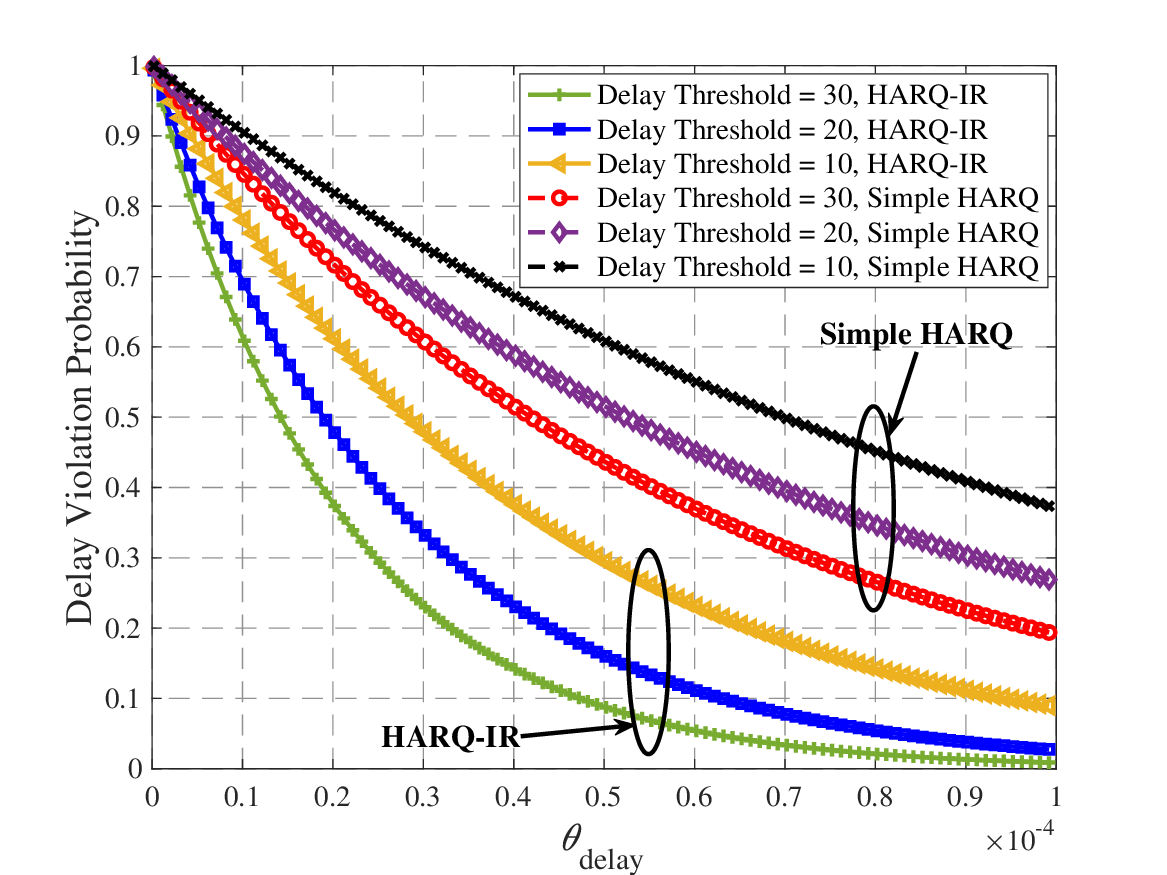}
		\caption{The delay violation probability vs. delay-bounded QoS exponent for our developed formulations using FBC.}
		\label{fig:06}
		\vspace{-8pt}
	\end{figure}

\begin{figure}[!t]
	\vspace{-5pt}
	\centering
	\includegraphics[scale=0.31]{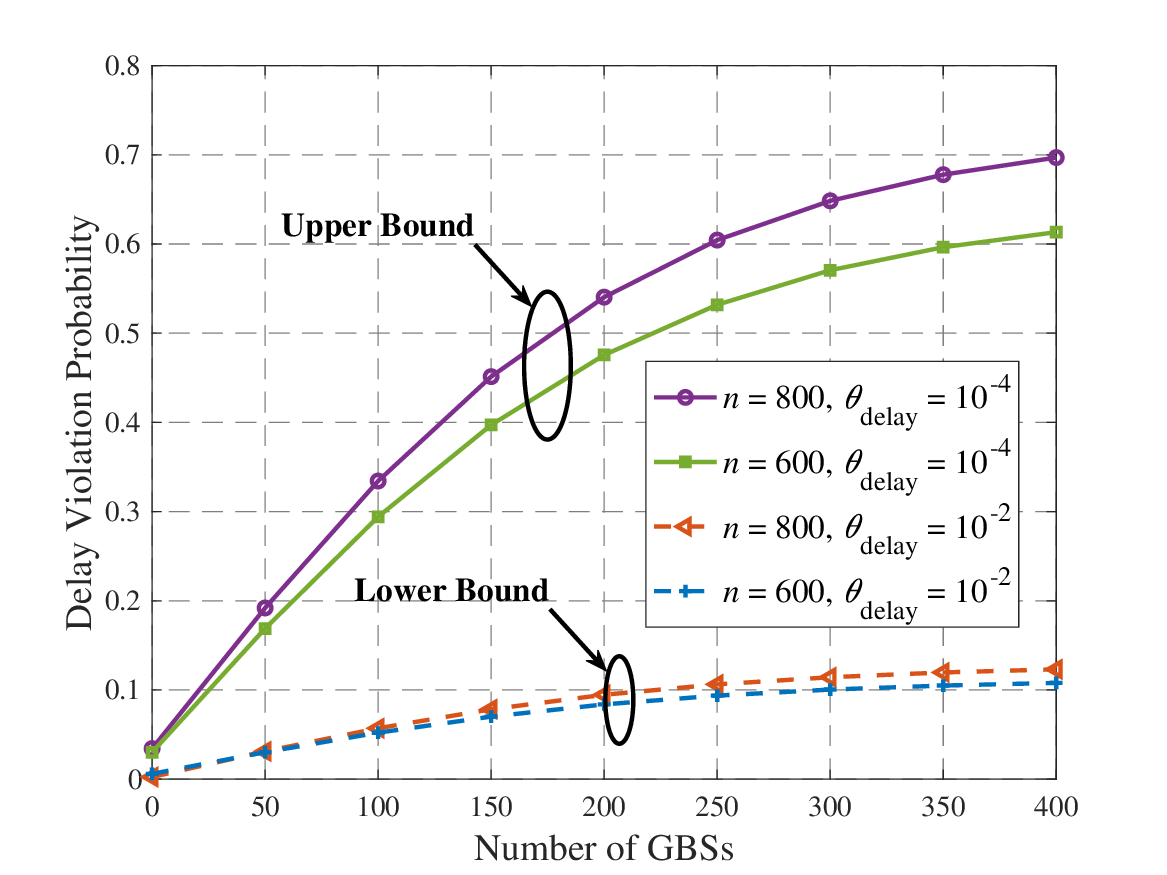}
	\caption{The delay violation probability vs. the number of GBSs for our developed formulations using FBC.}
	\label{fig:07}
	\vspace{-5pt}
\end{figure}

Figure~\ref{fig:06} presents the delay violation probability $p_{s}^{\text{delay}}(u)$ against the delay-bounded QoS exponent $\theta_{\text{delay}}$ for our developed formulations by applying HARQ-IR as compared with the simple HARQ scheme.
Fig.~\ref{fig:06} also shows that the HARQ-IR scheme outperforms the simple HARQ scheme in terms of the delay violation probability.
Furthermore, Fig.~\ref{fig:06} demonstrates that a larger $D_{\text{th},s}$ leads to a smaller delay violation probability.
In addition, Fig.~\ref{fig:07} plots the delay violation probability $p_{s}^{\text{delay}}(u)$ against the number of GBSs for our developed formulations.
Observing from Fig.~\ref{fig:06} and Fig.~\ref{fig:07}, it becomes evident that the delay violation probability $p_{s}^{\text{delay}}(u)$ declines as $\theta_{\text{delay}}$ increases. Consequently, a small delay-bounded QoS exponent establishes an upper bound, while a large delay-bounded QoS exponent sets a lower bound for the delay violation probability.
	
\begin{figure}[!t]		
			\vspace{-5pt}
		\centering
		\includegraphics[scale=0.31]{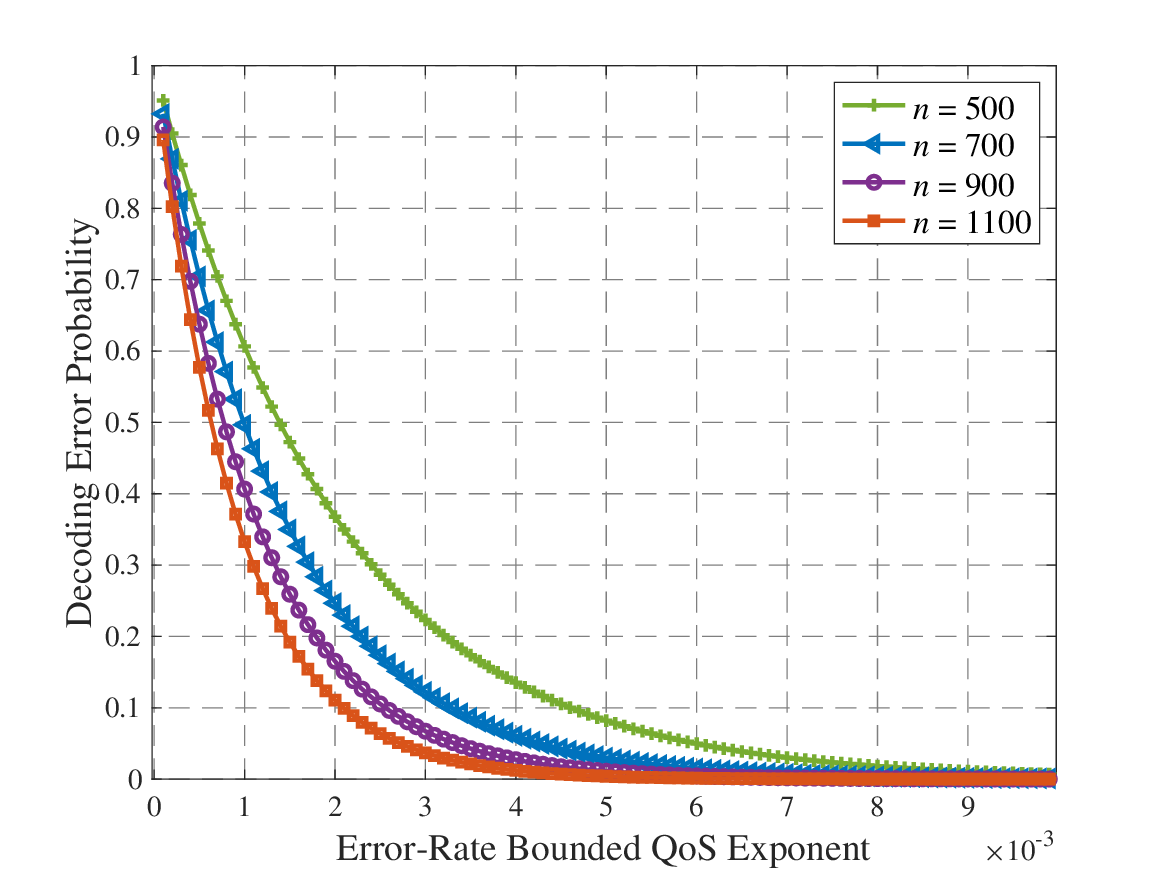}
		\caption{The decoding error probability $\epsilon_{s}(u)$ v.s. error-rate bounded QoS exponent $\theta_{\text{error}}$ for our developed formulations.}
		\label{fig:05a}
		\vspace{-5pt}
	\end{figure}

	Setting the average SNR =5 dB, Fig.~\ref{fig:05a} plots the behavior of the decoding error probability $\epsilon_{s}(u)$ concerning the error-rate bounded QoS exponent $\theta_{\text{error}}$ for the developed performance modeling formulations/frameworks.
	According to Fig.~\ref{fig:05a}, it becomes apparent that the decoding error probability exhibits an exponential decay pattern in relation to the error-rate bounded QoS exponent. This observation serves to validate the accuracy of Eq.~\eqref{equation039}.
	
\begin{figure}[!t]
		\vspace{-5pt}
		\centering
		\includegraphics[scale=0.31]{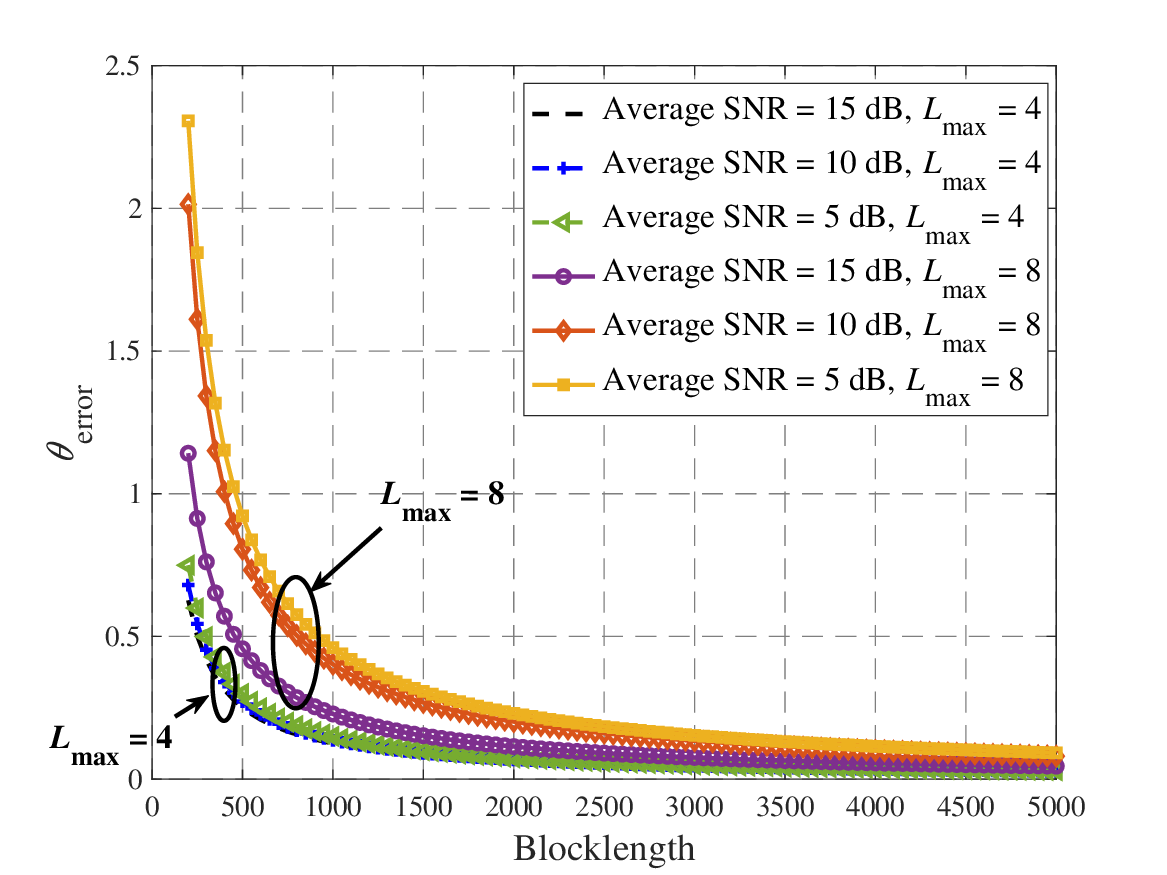}
		\caption{The error-rate bounded QoS exponent $\theta_{\text{error}}$ vs. the blocklength $n$.}
		\label{fig:05}
		\vspace{-5pt}
	\end{figure}

	Figure~\ref{fig:05} depicts the error-rate bounded QoS exponent with varying $n$ for our developed performance modeling formulations.
The graph depicted in Fig.~\ref{fig:05} showcases the exponential decay of the error-rate bounded QoS exponent as $n$ approaches infinity, thereby providing further confirmation of the definition of the error-rate bounded QoS exponent $\theta_{\text{error}}$ given in Eq.~\eqref{equation38}.
	Fig.~\ref{fig:05} also illustrates that when setting a large $L_{\max}$, there is an observable increase in the error-rate bounded QoS exponent. 
	This suggests the decoding error probability experiences a rapid decline as $L_{\max}$ increases, indicating an augmented level of reliability through applying the HARQ-IR protocol for our developed schemes.

	\section{Conclusions}\label{sec:conclusion}

We have developed a set of performance metrics for statistical QoS over 6G satellite-terrestrial integrated networks.
Particularly, first we have designed the satellite-terrestrial integrated wireless network architecture model and the AoI metric.
Second, we have characterized the peak-AoI bounded QoS metric and the corresponding violation probability functions using HARQ-IR protocol.
Third, we have introduced a novel collection of fundamental statistical QoS metrics, encompassing statistical QoS metrics in terms of delay and error-rate, leveraging FBC techniques.
Subsequently, a set of simulations have been performed to verify and assess the efficacy of our newly developed FBC-based modeling frameworks and analysis formulations in the context of mURLLC.

\begin{appendices}
	\section{Proof for Lemma 1}
		To obtain a tight upper-bound on the peak AoI violation probability, instead of the complex inversion formula, we apply the Mellin transform to formulate the moment bound as follows:
	\begin{align}\label{equation12b}
		p_{s}^{\text{AoI}}(u)&=\!\text{Pr}\left\{P_{s}^{\text{AoI}}(u)>e^{\frac{A_{\text{th}}}{n}}\right\}
		\!\leq\! e^{-\frac{\theta_{\text{AoI}} A_{\text{th}}}{n}}{\cal M}_{{\cal P}_{s}^{\text{AoI}}(u)}(1\!+\!\theta_{\text{AoI}})
		\nonumber\\
		&\leq e^{-\frac{\theta_{\text{AoI}} A_{\text{th}}}{n}}{\cal M}_{{\cal T}_{s}^{\text{I}}(u-1,u)}(1+\theta_{\text{AoI}}){\cal M}_{{\cal T}_{s}(u)}(1+\theta_{\text{AoI}})
	\end{align}
	for all $\theta_{\text{AoI}}>0$.
		Accordingly, based on Eq.~\eqref{equation005}, the total sojourn time	in the SNR domain is derived as follows:
		\begin{equation}
			{\cal T}_{s}(u)
			=\sup_{v\in\textsf{N}, v\leq u}\left\{\frac{T_{s}^{\text{S}}(v,u)}{T_{s}^{\text{I}}(v,u)}\right\}.
		\end{equation}
		The Mellin transform entails not only the computation of products and quotients, but also necessitates determining infimums and supremums.
		Correspondingly, we determine the Mellin transform of the sojourn time ${\cal M}_{{\cal T}_{s}(u)}(1+\theta_{\text{AoI}})$ as follows:
	\begin{align}\label{equation14}
			&{\cal M}_{{\cal T}_{s}(u)}(1+\theta_{\text{AoI}})
			=\mathbb{E}\left[\left(\sup_{v\in\textsf{N}, v\leq u}\left\{\frac{T_{s}^{\text{S}}(v,u)}{T_{s}^{\text{I}}(v,u)}\right\} \right)^{\theta_{\text{AoI}}}	\right]		\nonumber\\
			&\quad	=\mathbb{E}\left[\sup_{v\in\textsf{N}, v\leq u}\left\{\left(T_{s}^{\text{S}}(v,u)\right)^{\theta_{\text{AoI}}} \left(T_{s}^{\text{I}}(v,u) \right)^{-\theta_{\text{AoI}}}\right\}	\right]		\nonumber\\
			&\quad	\overset{(a)}{\leq}
			\sum_{v=1}^{u} {\cal M}_{{\cal T}_{s}^{\text{S}}(v,u)}(1+\theta_{\text{AoI}}){\cal M}_{{\cal T}_{s}^{\text{I}}(v,u)}(1-\theta_{\text{AoI}})
		\end{align}
		where ${\cal M}_{{\cal T}_{s}^{\text{S}}(v,u)}(\theta_{\text{AoI}})$ denotes the Mellin transform of the cumulative service time for the transmission of status-update $v$ up to $u$ in the SNR domain and ${\cal M}_{{\cal T}_{s}^{\text{I}}(v,u)}(\theta_{\text{AoI}})$ represents the Mellin transform of the inter-arrival time in the SNR domain.
		(a) in Eq.~\eqref{equation14} holds because of the non-negativity of ${\cal M}_{{\cal T}_{s}^{\text{S}}(v,u)}(1+\theta_{\text{AoI}})$ and ${\cal M}_{{\cal T}_{s}^{\text{I}}(v,u)}(1-\theta_{\text{AoI}})$.
		In addition, the union bound is employed to replace the supremum with a sum and their independence to evaluate the expectation of the products.
	Plugging Eq.~(\ref{equation14}) into Eq.~(\ref{equation12b}), we can obtain
	\begin{align}\label{equation16}
		p_{s}^{\text{AoI}}(u)\leq&\, e^{-\frac{\theta_{\text{AoI}} A_{\text{th}}}{n}}{\cal M}_{{\cal T}_{s}^{\text{I}}(u-1,u)}(1+\theta_{\text{AoI}})
		\nonumber\\
		&\times \Bigg[\sum_{v=1}^{u} {\cal M}_{{\cal T}_{s}^{\text{S}}(v,u)}(1+\theta_{\text{AoI}})
		{\cal M}_{{\cal T}_{s}^{\text{I}}(v,u)}(1-\theta_{\text{AoI}})\Bigg].
	\end{align}
	Then, we define the following kernel function, denoted by $\mathsf{K}_{s}(\theta_{\text{AoI}},u)$:
	\begin{align}\label{equation15}
		\mathsf{K}_{s}(\theta_{\text{AoI}},u)\triangleq&\,{\cal M}_{{\cal T}_{s}^{\text{I}}(u-1,u)}(1+\theta_{\text{AoI}})
		\Bigg[\sum_{v=1}^{u} {\cal M}_{{\cal T}_{s}^{\text{S}}(v,u)}(1+\theta_{\text{AoI}})\nonumber\\
		&\times
		{\cal M}_{{\cal T}_{s}^{\text{I}}(v,u)}(1-\theta_{\text{AoI}})\Bigg].
	\end{align}	
	Correspondingly, by substituting Eq.~(\ref{equation15}) into Eq.~(\ref{equation16}), Eq.~(\ref{equation026a}) is obtained, completing the proof of Lemma~1.
	
	\section{Proof for Theorem 1}
			We proceed with this proof by showing Theorem~\ref{theorem002a}'s \underline{Claim 1} and \underline{Claim 2}, respectively, as follows.
		
		\underline{Claim 1.} Based on Eqs.~(\ref{equation16}),~(\ref{equation18}),~(\ref{equation19}),  and~(\ref{equation18b}), the upper bound for peak AoI violation probability is rewritten as follows:
	\begin{align}\label{equation29}
		&p_{s}^{\text{AoI}}(u)
		\nonumber\\
		&\quad\!\!\!\leq\! e^{-\frac{\theta_{\text{AoI}} A_{\text{th}}}{n}}\!e^{\theta_{\text{AoI}}\left[\rho_{s}^{\text{I}}\left(\theta_{\text{AoI}}\right)+\sigma_{s}^{\text{I}}\left(\theta_{\text{AoI}}\right)\right]}
		e^{\theta_{\text{AoI}}\left[\sigma_{s}^{\text{I}}\left(\!-\theta_{\text{AoI}}\right)+\rho_{s}^{\text{S}}\left(\theta_{\text{AoI}}\right)+\sigma_{s}^{\text{S}}\left(\theta_{\text{AoI}}\right)\!\right]}
		\nonumber\\
		&\qquad\!\!\times\Bigg[\sum_{v=1}^{u} e^{-\theta_{\text{AoI}}\left[\rho_{s}^{\text{I}}\left(-\theta_{\text{AoI}}\right)-\rho_{s}^{\text{S}}\left(\theta_{\text{AoI}}\right)\right](u-v)}\Bigg]
		\nonumber\\
		&\quad\!\!\!	\overset{(a)}{=}\!
		\frac{e^{-\frac{\theta_{\text{AoI}} A_{\text{th}}}{n}}\!e^{\theta_{\text{AoI}}\left[\rho_{s}^{\text{I}}\left(\theta_{\text{AoI}}\right)+\sigma_{s}^{\text{I}}\left(\theta_{\text{AoI}}\right)\!\right]}
			e^{\theta_{\text{AoI}}\left[\sigma_{s}^{\text{I}}\left(\!-\theta_{\text{AoI}}\right)+\rho_{s}^{\text{S}}\left(\theta_{\text{AoI}}\right)+\sigma_{s}^{\text{S}}\left(\theta_{\text{AoI}}\right)\!\right]}}
		{1-e^{-\theta_{\text{AoI}}\left[\rho_{s}^{\text{I}}\left(-\theta_{\text{AoI}}\right)-\rho_{s}^{\text{S}}\left(\theta_{\text{AoI}}\right)\right]}}
	\end{align}
	where the following stability condition holds considering the case of G$|$G arrival and service processes:
	\begin{equation}\label{equation036}
		e^{-\theta_{\text{AoI}}\left[\rho_{s}^{\text{I}}\left(-\theta_{\text{AoI}}\right)-\rho_{s}^{\text{S}}\left(\theta_{\text{AoI}}\right)\right]}<1,
	\end{equation}
	leading to $\rho_{s}^{\text{S}}\left(\theta_{\text{AoI}}\right)<\rho_{s}^{\text{I}}\left(-\theta_{\text{AoI}}\right)$.
	(a) in Eq.~(\ref{equation29}) holds because we apply the sum of infinite geometric series for $\rho_{s}^{\text{S}}\left(\theta_{\text{AoI}}\right)<\rho_{s}^{\text{I}}\left(-\theta_{\text{AoI}}\right)$.
	Thus, plugging Eq.~\eqref{equation034} into Eq.~\eqref{equation036}, we can obtain Eq.~\eqref{equation17a}, completing the proof for \underline{Claim 1} of Theorem~\ref{theorem002a}.
		
		\underline{Claim 2.} As the MGF in terms of the sum of independent random variables $\mathcal{X}$ and $\mathcal{Y}$ is equal to the product of their individual MGFs, we can express this as ${\cal M}_{{\cal X+Y}}(1+\theta_{\text{AoI}})={\cal M}_{{\cal X}}(1+\theta_{\text{AoI}}){\cal M}_{{\cal Y}}(1+\theta_{\text{AoI}})$. 
	In the special case of GI$|$GI arrival and service processes,
	since ${\cal M}_{{\cal T}_{s}^{\text{I}}(v,u)}(1+\theta_{\text{AoI}})=\left[{\cal M}_{{\cal T}_{s}^{\text{I}}(u-1,u)}(1+\theta_{\text{AoI}})\right]^{(u-v)}$ and ${\cal M}_{{\cal T}_{s}^{\text{S}}(v,u)}(1+\theta_{\text{AoI}})
	=\left[{\cal M}_{{\cal T}_{s}^{\text{S}}(u)}(1+\theta_{\text{AoI}})\right]^{(u-v)}$, we can obtain the following minimal traffic and service parameters based on Eqs.~(\ref{equation18}) and~(\ref{equation18b}):
	\begin{equation}\label{equation29a}
		\begin{cases}
			\sigma_{s}^{\text{I}}\left(\theta_{\text{AoI}}\right)=\sigma_{s}^{\text{S}}\left(\theta_{\text{AoI}}\right)=0;\\
			\rho_{s}^{\text{I}}\left(-\theta_{\text{AoI}}\right)=-\frac{1}{\theta_{\text{AoI}}}\log\left\{\mathbb{E}\left[e^{-\theta_{\text{AoI}} T_{s}^{\text{I}}(u-1,u)}\right]\right\};\\
			\rho_{s}^{\text{S}}\left(\theta_{\text{AoI}}\right)=\frac{1}{\theta_{\text{AoI}}}\log\left\{\mathbb{E}\left[e^{\theta_{\text{AoI}} T_{s}^{\text{S}}(u)}\right]\right\}.
		\end{cases}
	\end{equation}
	By plugging the above traffic and service parameters into Eq.~(\ref{equation29}), we obtain
	\begin{align}\label{equation30}
		p_{s}^{\text{AoI}}(u)
		&\leq\frac{e^{-\frac{\theta_{\text{AoI}} A_{\text{th}}}{n}}{\cal M}_{{\cal T}_{s}^{\text{I}}(u-1,u)}(1+\theta_{\text{AoI}}){\cal M}_{{\cal T}_{s}^{\text{S}}(u)}(1+\theta_{\text{AoI}})}{1-{\cal M}_{{\cal T}_{s}^{\text{I}}(u-1,u)}(1-\theta_{\text{AoI}}){\cal M}_{{\cal T}_{s}^{\text{S}}(u)}(1+\theta_{\text{AoI}})},
	\end{align}
	while the stability condition holds, completing the proof for \underline{Claim 2}. Thus, we complete the proof of Theorem~\ref{theorem002a}.
		
			\section{Proof for Theorem 2}
			The proof is proceeded by establishing \underline{Claim 1} and \underline{Claim 2} successively.
		
		\underline{Claim 1.}  We prove \underline{Claim 1} through the following three steps, respectively.
		
		\underline{Step 1.}
		The decoding error probability $\epsilon_{s,l}(u)$ during the transmission of status-update $u$ is derived as follows:
		\begin{equation}\label{equation049}
			\epsilon_{s,l}(u)\approx \int_{0}^{\infty} {\cal Q}\left(\frac{\sqrt{n}\left[C(x)-R^{*}_{s,l}\right]}{\sqrt{V(x)}}\right) f_{\gamma_{s}(u)}\left(x\right) dx
		\end{equation}
		where $f_{\gamma_{s}(u)}\left(x\right)$ is the PDF of the SINR.
		Since $Q$-function has a complex form, it is difficult to find a closed-form expression for the decoding error probability. 
		Accordingly, we present an alternative approximation of the $Q$-function as follows:
		\begin{equation}\label{equation026}
			Q\left(\frac{C\left(\gamma_{s}(u)\right)-R^{*}_{s,l}}{\sqrt{V\left(\gamma_{s}(u)\right)/n}}\right)\approx \Psi(\gamma_{s}(u))
		\end{equation}
		where $\Psi(\gamma_{s}(u))$ is expressed as follows~\cite{BM2014}:
		\begin{equation}\label{equation027}
			\Psi\!(\gamma_{s}(u)\!)\!=\!
			\begin{cases}
				\!1, \qquad\qquad\qquad\qquad\qquad\quad \gamma_{s}(u)\!\leq\! \zeta_{\text{low},l}; \\
				\!\frac{1}{2}\!-\!\vartheta_{s,l}\sqrt{n}\!\left(\!\gamma_{s}(u)\!-\!2^{R^{*}_{s,l}-1}\!\right)\!,  \zeta_{\text{low},l}\!<\!\gamma_{s}(u)\!<\!\zeta_{\text{up},l}; \\
				\!0, \qquad\qquad\qquad\qquad\qquad\quad \gamma_{s}(u)\!\geq\! \zeta_{\text{up},l}.
			\end{cases}
		\end{equation}
		Taking expectation over Eqs.~(\ref{equation026}) and~(\ref{equation027}), we can obtain
		\begin{align}\label{equation028}
			\epsilon_{s,l}(u)
			&\approx
			F_{\gamma_{s}\!(u)}\left(\zeta_{\text{low},l}\right)\!+\!\!\Bigg[\frac{1}{2}\!+\!\vartheta_{s,l}\sqrt{n}\left(e^{R^{*}_{s,l}}\!-\!1\!\right)\!\!\Bigg]
			\!\Bigg[\!F_{\gamma_{s}(u)}\!\left(\zeta_{\text{up},l}\right)
			\nonumber\\
			& 	-F_{\gamma_{s}(u)}(\zeta_{\text{low},l})\Bigg]
			\!-\vartheta_{s,l}\sqrt{n}\int_{\zeta_{\text{low},l}}^{\zeta_{\text{up},l}}xf_{\gamma_{s}(u)}(x)dx
		\end{align}
		where $F_{\gamma_{s}(u)}(x)$ is the CDF of the SINR $\gamma_{s}(u)$.
		
		\underline{Step 2.} We assume that the power gain between GBS $j$ and the satellite decays exponentially with parameter $\alpha$ and follows Gamma distributions with a shape parameter $k_{\text{pg}}$ and a scale parameter $\eta_{\text{pg}}$.
The CDF of the channel gain $|h_{s}(u)|^{2}$, denoted by $F_{|h_{s}(u)|^{2}}(x)$, can be obtained as follows:
	\begin{equation}\label{equation02b}
		F_{|h_{s}(u)|^{2}}(x)=\alpha_{s}(u) \sum_{i=1}^{\infty}\frac{(m_{s}(u))_{i}[\delta_{s}(u)]^{i}}{(i!)^{2}[\beta_{s}(u)]^{i+1}}\gamma(i+1, \beta_{s}(u) x)
	\end{equation}
	where $\gamma(\cdot,\cdot)$ is the lower incomplete Gamma function.

To obtain a tractable model for the aggregate interference, we approximate the BS interference distribution using the Gamma model considering Rayleigh fading. 
By Campbell's theorem~\cite{6042301}, the mean aggregate interference is the same for all stationary point processes of the same intensity.
We derive the characteristic function of the aggregate interference, denoted by $\Phi_{I_{\text{a}}}$, as follows:
\begin{equation}\label{equation071}
	\Phi_{I_{\text{a}}}(\omega)\!=\!\exp\!\left\{\!\!-2\pi\lambda_{M}\!\!\!\!\!\!\int\limits_{h_{j}(u)}\!\!\int\limits_{\mathbb{R}}\!\left[1\!-\!e^{\jmath\omega x[d_{j}(u)]^{-\widetilde{\alpha}}}\right]\!dh_{j}(u)dd_{j}(u)\!\right\}
\end{equation}
where $\jmath=\sqrt{-1}$.
	Based on Eq.~\eqref{equation071}, we can obtain the closed-form expression of the $i^{\text{th}}$ cumulant of $\Phi_{I_{\text{a}}}(\omega)$ as follows:
	\begin{equation}\label{equation072}
		\kappa_{I_{\text{a}}}(i)=\frac{1}{j^i}\frac{d^{i}}{d \omega^{i} }\frac{(\log  \Phi_{I_{\text{a}}}(\omega))}{1}\Big|_{\omega=0}
	\end{equation}
By integrating Eq.~\eqref{equation071}, the result is obtained as follows~\cite{4453888}:
	\begin{equation}\label{equation078}
	\kappa_{I_{\text{a}}}(i)=\frac{2\pi\lambda_{M}}{i\widetilde{\alpha}-2}\mathbb{E}_{h_{j}(u)}\left[\left[h_{j}(u)\right]^{\frac{2}{\widetilde{\alpha}}}\right]
\end{equation}
	Accordingly, to obtain the closed form expressions of the PDF of aggregate interference power, denoted by $f_{I_{\text{a}}}(x)$, under the Gamma model, we approximately derive the PDF of the aggregate interference as follows:
	\begin{align}\label{equation77}
		f_{I_{\text{a}}}(x;k_{I_{\text{a}}},\eta_{I_{\text{a}}})
		&=\frac{1}{\Gamma(k_{I_{\text{a}}})(\eta_{I_{\text{a}}})^{k_{I_{\text{a}}}}}x^{k_{I_{\text{a}}}-1}\exp\left\{-\frac{x}{\eta_{I_{\text{a}}}}\right\}.
	\end{align}
	where
	\begin{equation}
	\begin{cases}
		k_{I_{\text{a}}}=\frac{\kappa_{I_{\text{a}}}(1)}{\kappa_{I_{\text{a}}}(2)}=\frac{\mathbb{E}\left[I_{\text{a}}\right]}{I\left[(I_{\text{a}})^{2}\right]};\\
		\eta_{I_{\text{a}}}=\frac{\kappa_{I_{\text{a}}}(2)}{\kappa_{I_{\text{a}}}(1)}=\frac{I\left[(I_{\text{a}})^{2}\right]}{\mathbb{E}\left[I_{\text{a}}\right]}.
	\end{cases}
	\end{equation}
Given the distribution of $I_{\text{a}}$, the CDF of the SINR is given:
		\begin{align}\label{equation053}
			F_{\gamma_{s}(u)}(x)&=\text{Pr}\left\{\frac{\phi_{s}(u){\cal P}_{s}(u)|h_{s}(u)|^{2}}{I_{\text{a}}+1} \leq x \Bigg|I_{\text{a}}\right\}
			\nonumber\\
			&=\int_{0}^{\infty}F_{|h_{s}(u)|^{2}}\left(\frac{x(y+1)}{\phi_{s}(u){\cal P}_{s}(u)}\right)f_{I_{\text{a}}}(y)dy.
		\end{align}
		By assuming that the interference dominates the noise{, i.e., $I_{\text{a}}\gg1$, we can rewrite the SINR as follows:
		\begin{equation}\label{equation080}
			\gamma_{s}(u)=\frac{\phi_{s}(u){\cal P}_{s}(u)|h_{s}(u)|^{2}}{I_{\text{a}}}.
		\end{equation}
		Accordingly, we can rewrite Eq.~\eqref{equation053} as follows:
		\begin{align}\label{equation081}
F_{\gamma_{s}(u)}\!(x)
=\int_{0}^{\infty}F_{|h_{s}(u)|^{2}}\left(\frac{xy}{\phi_{s}(u){\cal P}_{s}(u)}\right)f_{I_{\text{a}}}(y)dy.
\end{align}
		Then, plugging Eq.~\eqref{equation77} into Eq.~\eqref{equation080}, we have
		\begin{align}\label{equation055}
			F_{\gamma_{s}(u)}\!(x)\!
			&=\frac{\alpha_{s}(u)}{\Gamma(k_{I_{\text{a}}})(\eta_{I_{\text{a}}})^{k_{I_{\text{a}}}}} \sum_{i=1}^{\infty}\frac{(m_{s}(u))_{i}[\delta_{s}(u)]^{i}}{(i!)^{2}[\beta_{s}(u)]^{i+1}}
			 \nonumber\\
			&\quad\times\int_{0}^{\infty}\gamma\left(i+1,  \frac{\beta_{s}(u)xy}{\phi_{s}(u){\cal P}_{s}(u)}\right) y^{k_{I_{\text{a}}}-1}e^{-\frac{y}{\eta_{I_{\text{a}}}}}dy  \nonumber\\
			&=\frac{\alpha_{s}(u)}{\Gamma(k_{I_{\text{a}}})(\eta_{I_{\text{a}}})^{k_{I_{\text{a}}}}} \sum_{i=1}^{\infty}\frac{(m_{s}(u))_{i}[\delta_{s}(u)]^{i}}{(i!)^{2}[\beta_{s}(u)]^{i+1}} \nonumber\\
			&\quad\times\!\left[\frac{\beta_{s}(u)x}{\phi_{s}(u){\cal P}_{s}(u)}\right]^{i+1}\Gamma(i+1) \sum_{j=0}^{\infty}\frac{1}{\Gamma\left(i+j+2\right)} \nonumber\\
			&\quad\times\! \left[\!\frac{\beta_{s}(u)x}{\phi_{s}(u){\cal P}_{s}(u)}\!\right]^{j} \!\!\!\int_{0}^{\infty}\!\!\!\! y^{i+j+k_{I_{\text{a}}}}e^{-\left[\frac{\beta_{s}(u)x}{\phi_{s}(u){\cal P}_{s}(u)}+\frac{1}{\eta_{I_{\text{a}}}}\right]y}\!dy \nonumber\\
			&\!=\!\frac{\alpha_{s}(u)}{\Gamma(k_{I_{\text{a}}})(\eta_{I_{\text{a}}})^{k_{I_{\text{a}}}}} \!\!\sum_{i=1}^{\infty}\!\!\frac{(m_{s}(u)\!)_{i}[\delta_{s}(u)]^{i}}{(i!)^{2}[\beta_{s}(u)]^{i+1}}
			\!\!\left[\!\frac{\beta_{s}(u)x}{\phi_{s}(u){\cal P}_{s}(u)}\!\right]^{\!i+1}
			\nonumber\\
			&\quad\!\!\times\!\Gamma(i+1)\sum_{j=0}^{\infty}\frac{1}{\Gamma\left(i+j+2\right)}\left[\frac{\beta_{s}(u)x}{\phi_{s}(u){\cal P}_{s}(u)}\right]^{j}
			\nonumber\\
			&\quad\!\!\times\!
			\left[\frac{\beta_{s}(u)x}{\phi_{s}(u){\cal P}_{s}(u)}\!+\!\frac{1}{\eta_{I_{\text{a}}}}\right]^{-(i+j+k_{I_{\text{a}}}+1)}
			\!\!\!\!\!\!\!\!\!\!\Gamma\left(i\!+\!j\!+\!k_{I_{\text{a}}}\!+\!1\right).
		\end{align}}
		Then, the PDF of the SINR $\gamma_{s}(u)$ is derived as follows:
		\begin{align}\label{equation058}
			f_{\gamma_{s}(u)}\left(x\right)&=\int_{0}^{\infty}(y+1)f_{|h_{s}(u)|^{2}}\left(\frac{x(y+1)}{\phi_{s}(u){\cal P}_{s}(u)}\right)f_{I_{\text{a}}}(y)dy.
		\end{align}
		Similarly, by assuming that the interference dominates the noise, the PDF of the SINR $\gamma_{s}(u)$ is rewritten as follows:
		\begin{align}\label{equation059}
			&f_{\gamma_{s}(u)}\left(x\right)=\int_{0}^{\infty}yf_{|h_{s}(u)|^{2}}\left(\frac{xy}{\phi_{s}(u){\cal P}_{s}(u)}\right)f_{I_{\text{a}}}(y)dy
			\nonumber\\
			&=	\frac{\alpha_{s}(u)}{\Gamma(k_{I_{\text{a}}})(\eta_{I_{\text{a}}})^{k_{I_{\text{a}}}}}\!\!
			\sum_{l=0}^{m_{s}(u)-1}\!\frac{(-1)^{l}\left(1-m_{s}(u)\right)_{l}}{(l!)^{2}}
			\!\left[ \frac{\delta_{s}(u)x}{\phi_{s}(u){\cal P}_{s}(u)}\right]^{l}
			\nonumber\\
			&\quad \times \int_{0}^{\infty}
			y^{l+k_{I_{\text{a}}}}	e^{-\left\{\frac{\left[\beta_{s}(u)-\delta_{s}(u)\right] x}{\phi_{s}(u){\cal P}_{s}(u)}+\frac{1}{\eta_{I_{\text{a}}}}\right\}y}
			dy.
		\end{align}
		According to~\cite{LS2007}, we have
		\begin{align}\label{equation070}
			&f_{\gamma_{s}(u)}\left(x\right)=	\frac{\alpha_{s}(u)}{\Gamma(k_{I_{\text{a}}})(\eta_{I_{\text{a}}})^{k_{I_{\text{a}}}}}
			\sum_{l=0}^{m_{s}(u)-1}\frac{(-1)^{l}\left(1-m_{s}(u)\right)_{l}}{(l!)^{2}}
			\nonumber\\
			& \times\! \left[ \frac{\delta_{s}(u)x}{\phi_{s}(u){\cal P}_{s}(u)}\right]^{l}\!\!\! (l+k_{I_{\text{a}}})!\!
			\left\{\!\frac{\left[\beta_{s}(u)-\delta_{s}(u)\right] x}{\phi_{s}(u){\cal P}_{s}(u)}\!+\!\frac{1}{\eta_{I_{\text{a}}}}\!\right\}^{\!-l-k_{I_{\text{a}}}-1}\!\!.
		\end{align}
		To derive the decoding error probability, the integral term in Eq.~\eqref{equation028} need to be obtained.
		We define the auxiliary function $\Lambda_{l}(u)$ as follows:
		\begin{align}\label{equation073}
			\Lambda_{l}(u)\triangleq&\int_{\zeta_{\text{low},l}}^{\zeta_{\text{up},l}}xf_{\gamma_{s}(u)}(x)dx
			\nonumber\\
			=&
			\frac{\alpha_{s}(u)}{\Gamma(k_{I_{\text{a}}})(\eta_{I_{\text{a}}})^{k_{I_{\text{a}}}}}
			\sum_{l=0}^{m_{s}(u)-1}\frac{(-1)^{l}\left(1-m_{s}(u)\right)_{l}}{(l!)^{2}}
			\nonumber\\
			& \times \left[ \frac{\delta_{s}(u)}{\phi_{s}(u){\cal P}_{s}(u)}\right]^{l}	(l+k_{I_{\text{a}}})!\int_{\zeta_{\text{low},l}}^{\zeta_{\text{up},l}}x^{l+1}
			\nonumber\\
			& \times
			\left\{\frac{\left[\beta_{s}(u)-\delta_{s}(u)\right] x}{\phi_{s}(u){\cal P}_{s}(u)}+\frac{1}{\eta_{I_{\text{a}}}}\right\}^{-l-k_{I_{\text{a}}}-1}dx
		\end{align}
		which is Eq.~\eqref{equation073b}.
		Applying Eq.~(3.194) in~\cite{LS2007}, we obtain Eq.~\eqref{equation073b}.
		
		\underline{Step 3.} By substituting Eqs.~\eqref{equation055} and~\eqref{equation073} into Eq.~\eqref{equation028}, the decoding error probability is obtained as specified by Eq.~\eqref{theorem03_eq1}, concluding the proof of \underline{Claim 1} in Theorem~\ref{theorem03}.
		
		\underline{Claim 2.} The peak AoI violation probability after HARQ-IR retransmission round $l$ is derived as follows:
		\begin{align}\label{equation34}
			p_{s,l}^{\text{AoI}}(u)&\!\approx\!\frac{\lambda_{s} e^{-\frac{\theta_{\text{AoI}} A_{\text{th}}}{\widehat{n}}}}{\lambda_{s}-\theta_{\text{AoI}}}\exp\left\{ \theta_{\text{AoI}}\widehat{n}T\left[	1\!+\!\sum_{l=1}^{L-1}\epsilon_{s,l}(u)\right]\right\}, \,\,  \forall u
		\end{align}
		Then, plugging Eq.~\eqref{theorem03_eq1} into Eq.~\eqref{equation34}, the peak AoI violation probability after HARQ-IR retransmission $l$ is given in Eq.~\eqref{theorem03_eq2}, concluding the proof of \underline{Claim 2} in Theorem~\ref{theorem03}.
		Therefore, the proof of Theorem~\ref{theorem03} is concluded.
		
			\section{Proof for Lemma 2}
		Based on Eq.~\eqref{equation028}, the asymptotic decoding error probability $\epsilon^{\infty}_{s,l}(u)$ in the high SNR region is expressed as follows:
		\begin{align}\label{equation076}
			\epsilon^{\infty}_{s,l}(u)
			\approx&\,
			F^{\infty}_{\gamma_{s}(u)}\left(\zeta_{\text{low},l}\right)\!+\!\Bigg[\frac{1}{2}\!+\!\vartheta_{s,l}\sqrt{\widehat{n}}\left(e^{R^{*}_{s,l}}-1\right)\Bigg]
			\nonumber\\
			&\times \Bigg[F^{\infty}_{\gamma_{s}(u)}\left(\zeta_{\text{up},l}\right)\!-\!F^{\infty}_{\gamma_{s}(u)}(\zeta_{\text{low},l})\Bigg]
			\nonumber\\
			&-\!\vartheta_{s,l}\sqrt{\widehat{n}}\Bigg[\zeta_{\text{up},l}F^{\infty}_{\gamma_{s}(u)}(\zeta_{\text{up},l})-\zeta_{\text{low},l}F^{\infty}_{\gamma_{s}(u)}(\zeta_{\text{low},l})
			\nonumber\\
			&-\int_{\zeta_{\text{low},l}}^{\zeta_{\text{up},l}}F^{\infty}_{\gamma_{s}(u)}(x)dx\Bigg].
		\end{align}
		In addition, by substituting Eq.~\eqref{equation075} into Eq.~\eqref{equation076}, we have
		\begin{align}\label{equation077}
			\epsilon^{\infty}_{s,l}(u)
			\!\approx& \frac{\eta_{I_{\text{a}}}^{k_{I_{\text{a}}}+1}\Gamma\left(k_{I_{\text{a}}}+1\right)\alpha_{s}(u)\zeta_{\text{low},l}}
			{\Gamma(k_{I_{\text{a}}})(\eta_{I_{\text{a}}})^{k_{I_{\text{a}}}}\phi_{s}(u){\cal P}_{s}(u)}
			\!+\!\Bigg[\frac{1}{2}\!+\!\vartheta_{s,l}\sqrt{\widehat{n}}
			\nonumber\\
			&\!\times\! \left(e^{R^{*}_{s,l}}-1\right)\Bigg]
			\frac{\eta_{I_{\text{a}}}^{k_{I_{\text{a}}}+1}\Gamma\left(k_{I_{\text{a}}}+1\right)\alpha_{s}(u)}{\Gamma(k_{I_{\text{a}}})(\eta_{I_{\text{a}}})^{k_{I_{\text{a}}}}\phi_{s}(u){\cal P}_{s}(u)}
			\nonumber\\
			&\!\times\! \left(\zeta_{\text{up},l}\!-\!\zeta_{\text{low},l}\right)\!-\!\vartheta_{s,l}\sqrt{\widehat{n}}	\frac{\eta_{I_{\text{a}}}^{k_{I_{\text{a}}}+1}\Gamma\left(k_{I_{\text{a}}}+1\right)\alpha_{s}(u)}{\Gamma(k_{I_{\text{a}}})(\eta_{I_{\text{a}}})^{k_{I_{\text{a}}}}\phi_{s}(u){\cal P}_{s}(u)}
			\nonumber\\
			&\!\times\!\Bigg[\!\zeta_{\text{up},l}^{2}\!-\!\zeta_{\text{low},l}^{2}
			\!-\!\int_{\zeta_{\text{low},l}}^{\zeta_{\text{up},l}}\!\!\frac{\eta_{I_{\text{a}}}^{k_{I_{\text{a}}}+1}\Gamma\left(k_{I_{\text{a}}}+1\right)\alpha_{s}(u)x}
			{\Gamma(k_{I_{\text{a}}})(\eta_{I_{\text{a}}})^{k_{I_{\text{a}}}}\phi_{s}(u){\cal P}_{s}(u)}dx\!\Bigg]
			\nonumber\\
			=& \frac{\eta_{I_{\text{a}}}^{k_{I_{\text{a}}}+1}\Gamma\left(k_{I_{\text{a}}}+1\right)\alpha_{s}(u)}
			{\Gamma(k_{I_{\text{a}}})(\eta_{I_{\text{a}}})^{k_{I_{\text{a}}}}\phi_{s}(u){\cal P}_{s}(u)}
			\Bigg\{\zeta_{\text{low},l}\!+\!\Bigg[\frac{1}{2}\!+\!\vartheta_{s,l}\sqrt{\widehat{n}}
			\nonumber\\
			&\!\times\!\left(\!e^{R^{*}_{s,l}}\!-\!1\right)\!\!\Bigg]\!\! \left(\zeta_{\text{up},l}\!-\!\zeta_{\text{low},l}\right)\!-\!\vartheta_{s,l}\sqrt{\widehat{n}}\left(\zeta_{\text{up},l}^{2}-\zeta_{\text{low},l}^{2}\right)^{2}
			\nonumber\\
			& \!\times\!
			\left[1-\frac{\eta_{I_{\text{a}}}^{k_{I_{\text{a}}}+1}\Gamma\left(k_{I_{\text{a}}}+1\right)\alpha_{s}(u)}
			{2\Gamma(k_{I_{\text{a}}})(\eta_{I_{\text{a}}})^{k_{I_{\text{a}}}}\phi_{s}(u){\cal P}_{s}(u)}\right]
			\Bigg\}.
		\end{align}
		Thus, by substituting Eq.~(\ref{equation077}) in Eq.~(\ref{equation34}), we can obtain the asymptotic decoding error probability $p_{s,l}^{\text{AoI},\infty}(u)$ as specified by Eq.~\eqref{lemma02_eq01}.
		Therefore, the proof of Lemma~2 is concluded.			
					
					\section{Proof for Corollary 1}
		By applying Jensen's inequality, we can derive  $E_{0}\left[\rho_{s},P_{\bm{x}_{s}}\left(\bm{x}_{s}\right)\right]$ based on Eq.~\eqref{equation047} as follows:
		\begin{align}\label{equation068}
			&E_{0}\left[\rho_{s},P_{\bm{x}_{s}}\left(\bm{x}_{s}\right)\right]\!
			\geq -\frac{1}{n}\mathbb{E}_{\gamma_{s}(u)}\left[\log \left[\frac{\left(1+\rho_{s}+\gamma_{s}(u)\right)^{-n\rho_{s}}}
			{\left(1+\rho_{s}\right)^{-n\rho_{s}}}\right]\right]
			\nonumber\\
			&\quad=\!\rho_{s}\mathbb{E}_{\gamma_{s}(u)}\left[\log \left(1+\rho_{s}+\gamma_{s}(u)\right)\right]
			-\rho_{s}\log \left(1+\rho_{s}\right)
			\nonumber\\
			&\quad=\!\rho_{s}\mathbb{E}_{|h_{s}(u)|^{2},|h_{1}(u)|^{2},\cdots,|h_{K}(u)|^{2}}\Bigg[\log\Bigg[\phi_{s}(u){\cal P}_{s}(u)|h_{s}(u)|^{2}
			\nonumber\\
			&\qquad+(1+\rho_{s})\!\sum\limits_{j=1}^{K}\phi_{j} (u)
			P_{t}(u)|h_{j}(u)|^{2}+1+\rho_{s}\Bigg]\Bigg]		\nonumber\\
			&\qquad-\rho_{s}\mathbb{E}_{I_{\text{a}}}\left[\log\left({I_{\text{a}}+1}\right)\right]
			-\rho_{s}\log \left(1+\rho_{s}\right)
			\nonumber\\
			&\quad\!\approx\!\rho_{s}\!\log\!\Bigg\{\!\phi_{s}(u){\cal P}_{s}(u)\alpha_{s}(u)\!\!\left[\Gamma(2)\! \sideset{_2}{_{1}}{\mathop{F}}\!\left(m_{s}(u),2;1;\frac{\delta_{s}(u)}{\beta_{s}(u)}\right)\!\right]
			\nonumber\\
			&\qquad\times\! [\beta_{s}(u)]^{2} \! +\!(1\!+\!\rho_{s})\!\sqrt{\frac{\pi}{2}}\varsigma\!\sum\limits_{j=1}^{K}\!\phi_{j}(u)P_{t}(u)+\!1\!+\!\rho_{s}\Bigg\}
			\nonumber\\
			&\qquad-\rho_{s}\psi\left(k_{I_{\text{a}}}\right)-\rho_{s}\log\left(\eta_{I_{\text{a}}}\right)
			-\rho_{s}\log \left(1+\rho_{s}\right)
		\end{align}	
		which is Eq.~\eqref{theorem04_eq00}, concluding the proof of Corollary 1.						
								
\end{appendices}

	\nocite{*}
	\footnotesize
	\bibliographystyle{IEEEtran}
	\bibliography{myref.bib}

\end{document}